\documentclass[pre,preprint,showkeys,preprintnumbers]{revtex4}

\usepackage{graphicx}
\usepackage{dcolumn}
\usepackage{bm}
\usepackage{amsmath}
\usepackage{subfigure}
\usepackage{ulem}
\usepackage[dvips]{color}

\begin{document}

\preprint{Aubry et al.}

\title[Statistics of the propagation operator in random media]{Singular value distribution of the propagation matrix in random scattering media}

\author{Alexandre Aubry}
\affiliation{Institut Langevin, ESPCI ParisTech\\
CNRS UMR 7587, Universit\'e Denis Diderot (Paris VII),\\
Laboratoire Ondes et Acoustique, 10 rue Vauquelin, 75005 Paris, France}

\author{Arnaud Derode}
\affiliation{Institut Langevin, ESPCI ParisTech\\
CNRS UMR 7587, Universit\'e Denis Diderot (Paris VII),\\
Laboratoire Ondes et Acoustique, 10 rue Vauquelin, 75005 Paris, France}


\begin{abstract}
The distribution of singular values of the propagation operator in a random medium is investigated, in a backscattering configuration. Experiments are carried out with pulsed ultrasonic waves around 3 MHz, using an array of 64 programmable transducers placed in front of a random scattering medium. The impulse responses between each pair of transducers are measured and form the response matrix. The evolution of its singular values with time and frequency is computed by means of a short-time Fourier analysis.  The mean distribution of singular values exhibits very different behaviours in the single and multiple scattering regimes. The results are compared with random matrix theory. Once the experimental matrix coefficients are renormalized, experimental results and theoretical predictions are found to be in a very good agreement. Two kinds of random media have been investigated: a highly scattering medium in which multiple scattering predominates and a weakly scattering medium. In both cases, residual correlations that may exist between matrix elements are shown to be a key parameter. Finally, the possibility of detecting a target embedded in a random scattering medium based on the statistical properties of the strongest singular value is discussed.

\end{abstract}


\maketitle

\copyright{Taylor and Francis, 2010. This is the author's version of the work. It is posted here by permission of Taylor and Francis for personal use, not for redistribution. The definitive version was published in Waves in Random and Complex Media, Volume 20 Issue 3, August 2010, pages 333 - 363.
doi:10.1080/17455030903499698 (\href{http://dx.doi.org/10.1080/17455030903499698}{http://dx.doi.org/10.1080/17455030903499698})}

\section{Introduction}

Wave propagation in a multiple scattering environment has been an interdisciplinary subject of interest in a huge variety of domains ranging, \textit{e.g.}, from solid state physics to optics, electromagnetism or seismology since multiple scattering can occur with all kinds of waves, whether quantum or classical. Among all areas of mesoscopic wave physics, some (like acoustics, seismology, microwaves) have the experimental advantage to offer controllable multi-element arrays of quasi-pointlike emitters/receivers. In such a  case, the propagation between two arrays is best described by a matrix, termed the propagation operator $\mathbf{K}$. At each frequency, its coefficients $k_{ij}$ correspond to the complex response between array elements $i$ and $j$. Despite their diversity all practical applications of wave physics (communication, detection, imaging, characterization...), have one thing in common: all the available information is contained in the array response matrix $\mathbf{K}$. Once $\mathbf{K}$ is known, the rest is only post-processing. Therefore, in a random scattering environment it is essential to study the statistical properties of $\mathbf{K}$, and their relation to field correlations, weak localisation, single versus multiple scattering.

Previous works have been performed in a transmission context, whether it be for communication purposes \cite{tulino,sprik,nuah,moustakas2} or scattering problems \cite{vellekoop,pendry}. In this paper, we will consider backscattering configurations : the same array of $N$ independent elements is used to transmit and receive waves. In that case,  $\mathbf{K}$ is a square matrix of dimension $N\times N$, and it is symmetric if the medium is reciprocal. We are particularly interested in the singular value decomposition (SVD) of the propagation operator, which amounts to write $\mathbf{K}$ as the product of three matrices: $\mathbf{K}= \mathbf{U}\mathbf{\Lambda}\mathbf{V^{\dag}}$. $\mathbf{\Lambda}$ is a diagonal matrix whose nonzero elements $\lambda_i$ are called the singular values of $\mathbf{K}$. They are always real and positive, and arranged in a decreasing order $(\lambda_1>\lambda_2>...>\lambda_N)$. $\mathbf{U}$ and $\mathbf{V}$ are unitary matrices whose columns correspond to the singular vectors.

It is now well known that in the case of point-like scatterers in a homogeneous medium, each scatterer is mainly associated to one non-zero singular value of $\mathbf{K}$ \cite{prada,prada2}, as long as the number of scatterers is smaller than $N$ and multiple scattering is neglected \cite{prada4,minonzio}. So a singular value decomposition of $\mathbf{K}$ (or equivalently a diagonalisation of the so-called time-reversal operator $\mathbf{KK^*}$) allows the selective detection of several targets, each being associated to a singular value of $\mathbf{K}$. This is the core of a detection method named DORT (French acronym for Decomposition of the Time Reversal Operator) \cite{prada,prada2}. DORT has shown its efficiency in detecting and separating the responses of several scatterers in homogeneous or weakly heterogeneous media \cite{prada,borcea} as well as in waveguides \cite{mordant,lingevitch,Carin04}. It has found applications in non-destructive evaluation\cite{kerbrat}, underwater acoustics \cite{gaumond,prada3}, electromagnetism \cite{Saillard99,Micolau03,iakoleva2,iakoleva} and in radar applied to forest environments \cite{badereau,nguyen,tabbara}.

In this paper we will deal with random scattering media, consisting of a large number ($>>N$) of randomly distributed scatterers, showing possibly multiple scattering between them. We will also address the issue of detecting a stronger reflector hidden in a statistically homogenous scattering medium, based on the strongest singular value $\lambda_1$. Since the propagation medium is considered as one realisation of a random process, some general results of random matrix theory (RMT) may be fruitfully applied.

RMT has been widely used in physics, statistics and engineering. The domain of applications are numerous, ranging from nuclear physics \cite{brody} or chaotic systems \cite{ellegard} to neural networks \cite{cun}, telecommunications \cite{tulino} or financial analysis \cite{laloux}. RMT predicts general behaviours of stochastic systems as, for instance, determining the Shannon capacity for MIMO communications in random media \cite{nuah,moustakas2} or statistical properties of highly excited energy levels for heavy nuclei \cite{brody}. Another direct application of RMT is to separate the deterministic and random contributions in multivariate data analysis \cite{sengupta,johnstone,karoui}. 

Here, the experimental configuration we consider uses a piezo-electric array, with a finite ($N=$64) numbers of elements sending wide-band ultrasonic waves around 3 MHz in an a priori unknown scattering medium. The main issues we address in this work are the applicability of  RMT \cite{tulino} to this experimental context, and its interest to establish a detection criterion based on the statistical properties of $\lambda_1$. Since $\mathbf{K}$ is random, the relevant observable is the probability distribution function $\rho(\lambda)$ of its singular values. Recent experiments \cite{aubry09} indicate that in a multiple scattering regime, the distribution of the singular values is in good agreement with a simple law derived from RMT, the so-called ``quarter circle law'' $\rho_{QC}(\lambda)=\sqrt{4-\lambda^2}/\pi$ ($0<\lambda<2$) \cite{marcenko,tulino}. In theory, the quarter circle law applies to square matrices of infinite dimensions, containing independently and identically distributed elements, with zero-mean and variance $1/N$. 

Yet, from a physical point of view some of these assumptions do not hold: in a backscattering configuration the matrix elements are neither independent nor identically distributed, for several reasons (among others, reciprocity and weak localisation). In particular, the field-field correlations that may exist between matrix elements is a key parameter \cite{sengupta,nuah,moustakas2,sprik}. Moreover when single scattering dominates, the spectral behaviour of $\mathbf{K}$ is shown to be similar to that of a random Hankel matrix (\textit{i.e}, whose elements are constant along each antidiagonal), and not a ``classical'' random matrix. This is due to the persistence of a deterministic coherence between single-scattered signals along the antidiagonals of the matrix $\mathbf{K}$ despite randomness, which does not occur in the multiple-scattering regime \cite{aubry09}. Therefore in a practical situation, the probability distribution function $\rho(\lambda)$ may differ significantly from the simple quarter circle law. Finally, in the last part of this paper we present theoretical considerations about the detection of a target embedded in a random scattering medium. We introduce a detection criterion based on the statistical properties of the strongest singular value $\lambda_1$. This approach can be generalized to target detection in noisy environments.

\section{\label{sec:exp_num}Experimental procedure}

\begin{figure}[htbp] 
\begin{center}
\includegraphics{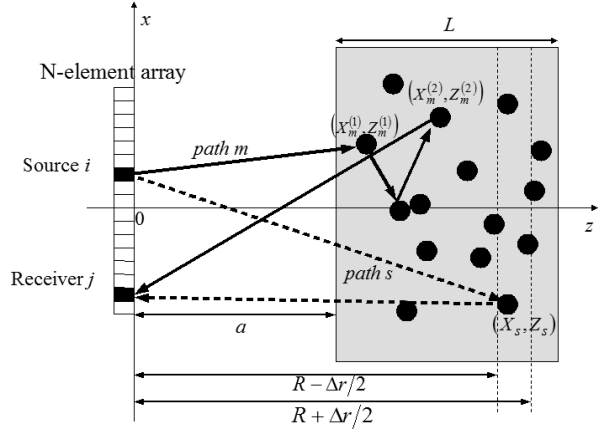}
\caption{\label{fig:fig1} Experimental setup. A 64-element array is placed in front of a random medium at
a distance $a$. The whole setup is immersed in a water tank. The inter-element response $k_{ij}(T,f)$, around the time of flight $T$ and at the frequency $f$, is measured. It contains contributions of single and multiple scattering paths whose lengths belong to the interval $[R-\Delta r/2;R+\Delta r/2]$, where $R=cT/2$ and $\Delta r = c\Delta t /2$. Examples of a single scattering path (labelled $s$, dashed line) and of a multiple-scattering path (labelled $m$, continuous line) is drawn. $(X_s,Z_s)$ are the coordinates of the scatterer involved in path $s$. $(X^{(1)}_m,Z^{(1)}_m)$ and $(X^{(2)}_m,Z^{(2)}_m)$ are the coordinates the first and last scatterers along path $m$.}
\end{center}
\end{figure}

\begin{figure}[htbp] 
\begin{center} 
\includegraphics{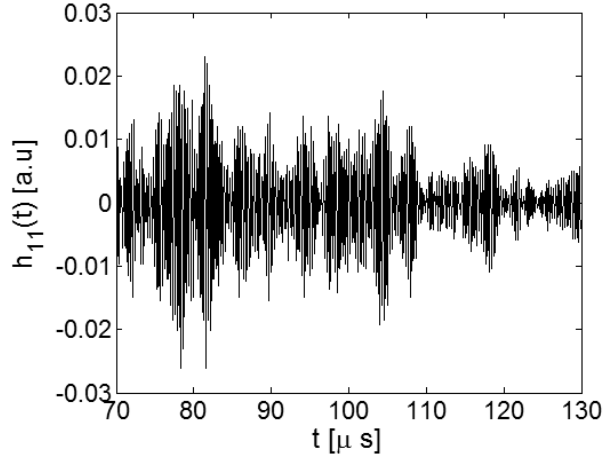}
\caption{\label{fig:fig1_2} Impulse response $h_{11}(t)$ measured in the forest of rods (see Sec.\ref{subsec:exp_obs_ms}).}
\end{center} 
\end{figure}

The experiment takes place in a water tank. We use a N-element ultrasonic array ($N=64$) with a 3 MHz central frequency and a 2.5-3.5 MHz bandwidth; each array element is 0.39 mm in size and the array pitch $p$ is 0.417 mm. The sampling frequency is 20 MHz. The array is placed in front of the medium of interest. The first step of the experiment consists in measuring the inter-element matrix (see Fig.~\ref{fig:fig1}). A $100$-$\mu$s-long linear chirp is emitted from transducer $i$ into the scattering sample immersed in water. The backscattered wave is recorded with the $N$ transducers of the same array. The operation is repeated for the $N$ emitting transducers. The response from transducer $i$ to transducer $j$ is correlated with the emitted chirp, which gives the impulse response $h_{ij}(t)$. A typical impulse response is shown in Fig.\ref{fig:fig1_2}. The $N \times N$ response matrix $\mathbf{H}(t)$ whose elements are the $N^2$ impulse responses $h_{ij}(t)$ is obtained. Because of reciprocity, $h_{ij}(t)=h_{ji}(t)$ and $\mathbf{H}(t)$ is symmetric. We take as the origin of time ($t=0$) the instant when the source emits the incident wave.

A scattering medium is essentially characterized by its scattering mean-free path $l_e$, and its diffusion constant $D$. If the scattering path length within the medium is larger than $l_e$, multiple scattering can predominate. This is expected to happen at late times in the scattered signals $h_{ij}(t)$. 

The impulse response matrix $\mathbf{H}(t)$ is truncated into short time windows in order to keep the temporal resolution provided by acoustical measurements and study the transition from a single scattering to a multiple scattering regime.
The time signals $h_{ij}(t)$ are truncated into overlapping windows : $k_{ij}(T,t)=h_{ij}(T-t)W_R(t)$ with $W_R(t)=1 \; \text{for} \; t\in[-\Delta t/2 \; , \; \Delta t/2]$, $W_R(t)=0$ elsewhere. The value of $\Delta t$ is chosen so that signals associated with the same scattering path within the medium arrive in the same time window. The detailed calculation of $\Delta t$ is given in Appendix \ref{app:temp_window}. In our experiments, we have typically $\Delta t \sim 30$ periods. At each time $T$, the $k_{ij}$ form a matrix $\mathbf{K}$. A short-time Fourier analysis is achieved by a discrete Fourier transform (DFT) and gives the response matrices $\mathbf{K}(T,f)$ at time $T$ and frequency $f$. The numerical SVD of each matrix $\mathbf{K}$ is performed and yields $N$ singular values $\lambda_i(T,f)$ at each time $T$ and frequency $f$. 

The next step consists in studying the distribution of singular values. It should be noted that we only have access to one realisation of disorder: the scattering medium is fixed, there is no ensemble averaging. Experimentally, the ensemble average can only be estimated by an average over frequency, and/or an average over time. The universal results provided by RMT are based upon the assumption that the elements of the random matrix have zero mean and a variance of $1/N$. The first condition is easily met, assuming that $k_{ij}(T,f)$ is a superposition of scattering contributions with a phase uniformly distributed between $-\pi$ and $+\pi$. In order to fulfill the second condition and compare experimental and theoretical results, $\mathbf{K}$ is renormalized into $\mathbf{\tilde{K}}$ with normalized singular values $\tilde{\lambda}_i$
\begin{equation}
\label{eqn:eq2}
\tilde{\lambda}_i=\frac{\lambda_i}{\sqrt{\frac{1}{N}\sum_{p=1}^N \lambda_p^2}}
\end{equation}

Once this renormalization is achieved, the experimental singular values distribution can be investigated. First, we form the histogram $\mathcal{H}(\lambda)$ of the whole set of renormalized singular values $\tilde{\lambda}_i (T,f)$, taken at every rank $i$, time $T$ and frequency $f$. Bins of this histogram are the intervals $[mw;(m+1)w]$, where $w$ is the bin width and $m$ is a non-negative integer. $\mathcal{H}(\lambda)$ denotes the number of singular values $\tilde{\lambda}_i (T,f)$ contained in the same bin as $\lambda$. An estimator of the probability density function of the singular values is obtained:
\begin{equation}
\label{eqn:eq3}
\hat{\rho}(\lambda)=\frac{\mathcal{H}(\lambda)}{nw}
\end{equation}
$n$ is the total number of singular values ($n=N \times n_T \times n_f$, $n_T$ is the number of time windows considered and $n_f$ the number of frequencies at which the DFT of $k_{ij}$ is achieved). At early times ($cT -2a <l_e$, with $c$ the wave speed in the surrounding medium) multiple scattering can be neglected, whereas at later times ($cT-2a>>l_e$) multiple scattering dominates. In the following of the study, the theoretical singular values distribution predicted by RMT will be confronted to the experimental estimator $\hat{\rho}$, both in multiple and single scattering regimes.

\section{\label{sec:ms}Multiple-scattering regime}

\subsection{\label{subsec:exp_obs_ms}Experimental configuration}
Here, we study the multiple-scattering regime. To that aim, we use a random-scattering slab consisting of parallel steel rods (longitudinal sound velocity $c_L=5.7$ mm/$\mu$s, transverse sound velocity $c_T=3$ mm/$\mu$s, radius 0.4 mm, density 7.85 kg/L) randomly distributed with a concentration $n$=12 rods/cm$^2$. The frequency-averaged elastic mean-free path $l_e$ is $7.7 \pm 0.3$ mm between 2.5 and 3.5 MHz \cite{derode4}. The distance $a$ between the array and the scattering sample is 25 mm. The slab thickness $L$ is 40 mm. Even in such strongly diffusive media, single scattering occurs for small times of flight. Here, we will consider times of flight $T$ larger than $70$ $\mu$s, corresponding to scattering path lengths more than 7 mean free paths. Under these conditions, single scattering can be neglected and $\mathbf{K}$ only contains multiple scattering contributions. The whole experimental procedure described in Sec.\ref{sec:exp_num} is achieved and a set of renormalized matrices $\mathbf{\tilde{K}}(T,f)$ is obtained. The time-window length $\Delta t$ is directly deduced from the calculation shown in Appendix \ref{app:temp_window} and is set to 10 $\mu$s. 

\subsection{\label{subsubsec:exp_obs_ms}Experimental distribution of singular values}
We perform singular value decompositions of $\mathbf{\tilde{K}}(T,f)$ and obtain the estimator $\hat{\rho}(\lambda)$ as defined in Eq.\ref{eqn:eq3}. According to RMT, if the matrix coefficients $\tilde{k}_{ij}$ are complex random variables independently and identically distributed with zero mean and a variance of $1/N$, then the asymptotic (\textit{i.e} for $N\rightarrow \infty$) density function of its singular values is given by the quarter circle law \cite{marcenko,tulino}.
\begin{figure}[htbp]
 \begin{center} 
\includegraphics{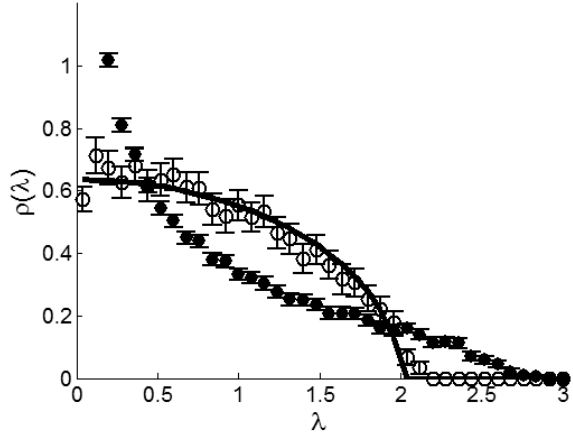}
\caption{\label{fig:fig2}
Experimental distributions of singular values of $\mathbf{\tilde{K}}$ (black disks) and $\mathbf{\tilde{K}_t}$ (white disks) are compared to the quarter circle law (black continuous line). The error bars are $\pm$ two standard deviations.}
\end{center} 
\end{figure}
As we can see in Fig.\ref{fig:fig2}, the experimental distribution of singular values is very far from this prediction. The major reason for this discrepancy is that the coefficients $\tilde{k}_{ij}$ are not independent. The experimental matrix $\mathbf{\tilde{K}}$ exhibits strong correlations between neighbour entries. They are measured by the correlation coefficient $\Gamma_m$
\begin{equation}
\label{eqn:corr_coeff}
\Gamma_m=\frac{\left < \tilde{k}_{i,j}\tilde{k}_{i,j+m}^*\right >_{T,f,(i,j)}}{\left < \left | \tilde{k}_{i,j}\right |^2 \right >_{T,f,(i,j)}}=\frac{\left < \tilde{k}_{i,j}\tilde{k}_{i+m,j}^*\right >_{T,f,(i,j)}}{\left < \left | \tilde{k}_{i,j}\right |^2 \right >_{T,f,(i,j)}}
\end{equation}
where the symbol $<. >$ denotes an average over the variables in the subscript, \textit{i.e} time $T$, frequency $f$ and source/receiver pairs $(i,j)$. The integer $m=i-j$ represents the distance between sources or receivers, in units of $p$ (the array pitch). Fig.\ref{fig:fig3}(a) clearly points out a strong correlation between neighbour entries, with a coefficient $\Gamma_1=\Gamma_{-1}\simeq 0.5$. The physical origin of these correlations will be detailed in Sec.\ref{subsec:correlation}. We will also show how to incorporate these correlations into the theoretical model for $\rho(\lambda)$.
\begin{figure}[htbp] 
\begin{center}
\begin{minipage}{140mm}
\subfigure[]{
\resizebox*{7cm}{!}{\includegraphics{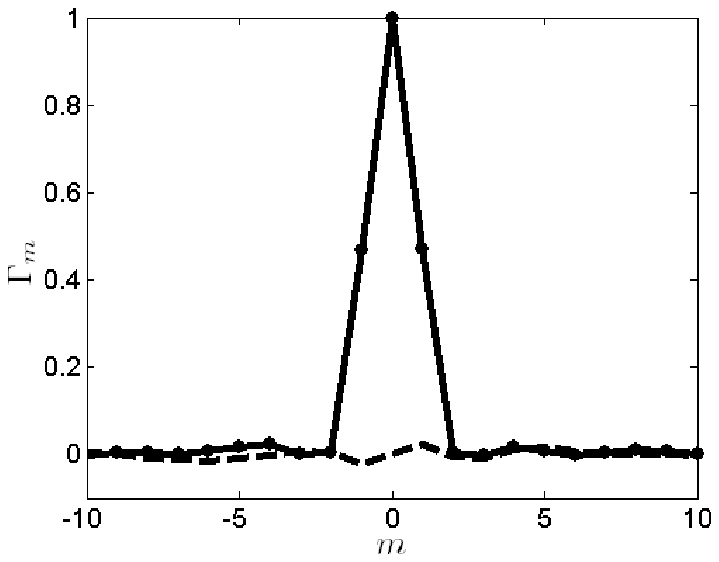}}}%
\subfigure[]{
\resizebox*{7cm}{!}{\includegraphics{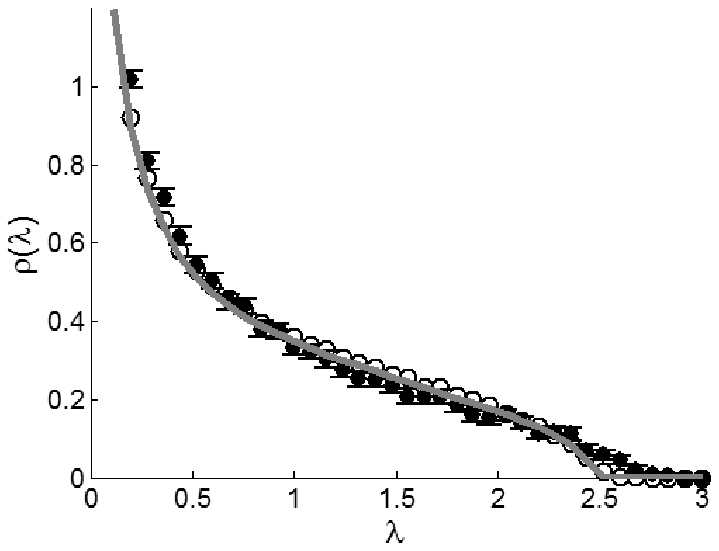}}}%
\caption{\label{fig:fig3}(a) Normalized correlations $\Gamma_m$ (real part : continous line; imaginary part : dashed line) estimated from experimental results. (b) The experimental distribution of singular values (black disks) is compared to the theory taking into account correlations (grey continuous line) and to the result of the numerical simulation (white disks) described in Sec.\ref{subsec:correlation}. The error bars are $\pm$ two standard deviations.}
\end{minipage}
\end{center}
\end{figure}

In this particular case, correlations can be simply removed by considering only one in two elements. From the initial matrix $\mathbf{K}$ of dimension $N \times N$, a truncated matrix $\mathbf{K}_t$ of dimension $N/2 \times N/2$ is built, considering only elements of odd or even index. $\mathbf{K}_t$ no longer exhibits short range correlations at emission and reception. The experimental distribution of normalized singular values is now much closer to the quarter circle law (see Fig.\ref{fig:fig2}). Nevertheless, a slight disagreement remains between the theoretical and experimental curves, specially at the vinicity of $\lambda=0$ and $\lambda=2$. 

In the next paragraph, we show that in the multiple scattering regime the variance of the matrix entries $k_{ij}$ is not the same for all pairs $(i,j)$, contrary to the assumption of identically distributed random variables. The consequence of this deviation from RMT assumptions will be discussed.

\subsection{\label{subsec:identical} Variance of coefficients $\tilde{k}_{ij}$.}

The signals $k_{ij}(T,f)$ at a time $T$ and frequency $f$ correspond to the sum of partial waves that reach the array in the time window $[T-\Delta t/2 ;T+\Delta t/2]$. They are associated with multiple scattering paths whose length belongs to the interval $[R-\Delta r/2;R+\Delta r/2]$, with $R=cT/2$ and $\Delta r=c\Delta t/2$. An example of such paths is drawn in Fig.\ref{fig:fig1}. The response $k_{ij}(T,f)$ can be decomposed into a sum of partial waves associated with the $N_q$ paths. In a 2D configuration, under the paraxial approximation and assuming point-like transducers and scatterers, $k_{ij}(T,f)$ can be expressed as
\begin{equation}
\label{eqn:eq_ms_signal}
k_{ij}(T,f) \propto \sum_{q=1}^{N_q} B_q   \frac{\exp \left [ j k \left ( Z^{(1)}_q + Z^{(2)}_q \right ) \right ]}{\sqrt{Z^{(1)}_qZ^{(2)}_q}} \exp \left[ j k \frac{ \left (x_i -X_q^{(1)} \right )^2}{2Z^{(1)}_q} \right ] \exp \left [ j k \frac{ \left (x_j -X_q^{(2)} \right )^2}{2Z^{(2)}_q} \right ]
\end{equation}
where $k=2 \pi f / c$ is the wave number in the surrounding medium. The index $q$ denotes the $q^{th}$ path which contributes to the signal received at time $T$. $\left (X_q^{(1)},Z_q^{(1)}\right )$ and $\left (X_q^{(2)},Z_q^{(2)}\right )$ are respectively the coordinates of the first and last scatterers along the path $q$. $B_q$ is the complex amplitude associated with path $q$, from the first scattering event at $\left (X_q^{(1)},Z_q^{(1)}\right )$ until the last one at $\left (X_q^{(2)},Z_q^{(2)}\right )$.
From the central limit theorem, we expect that the coefficients $k_{ij}$ are gaussian complex random variables. 

The renormalization (Eq.\ref{eqn:eq2}) leads to a set of matrices $\mathbf{\tilde{K}}$. If the $k_{ij}$ were identically distributed, the renormalized coefficients $\tilde{k}_{ij}$ would be complex random variables with variance $ \sigma_{ij}^2=1/N$.
\begin{figure}[htbp] 
\begin{center}
\begin{minipage}{140mm}
\subfigure[]{
\resizebox*{7cm}{!}{\includegraphics{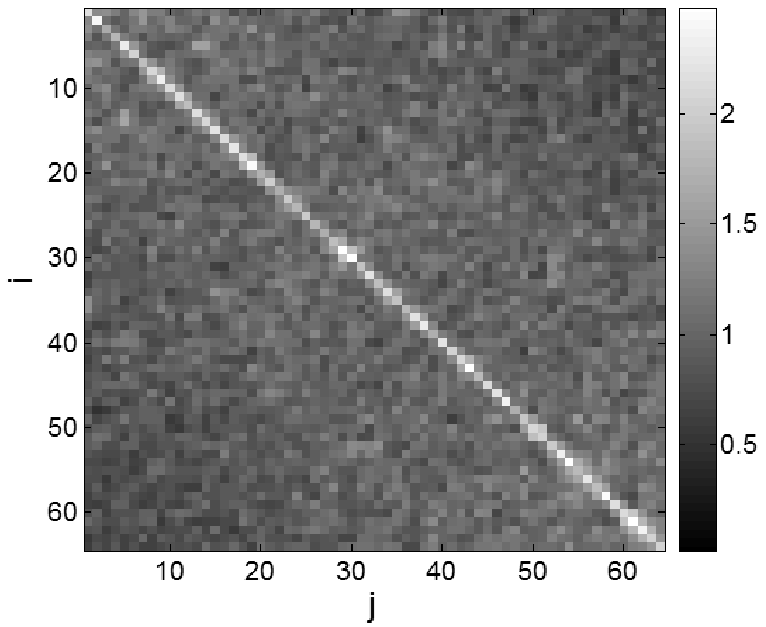}}}%
\subfigure[]{
\resizebox*{7cm}{!}{\includegraphics{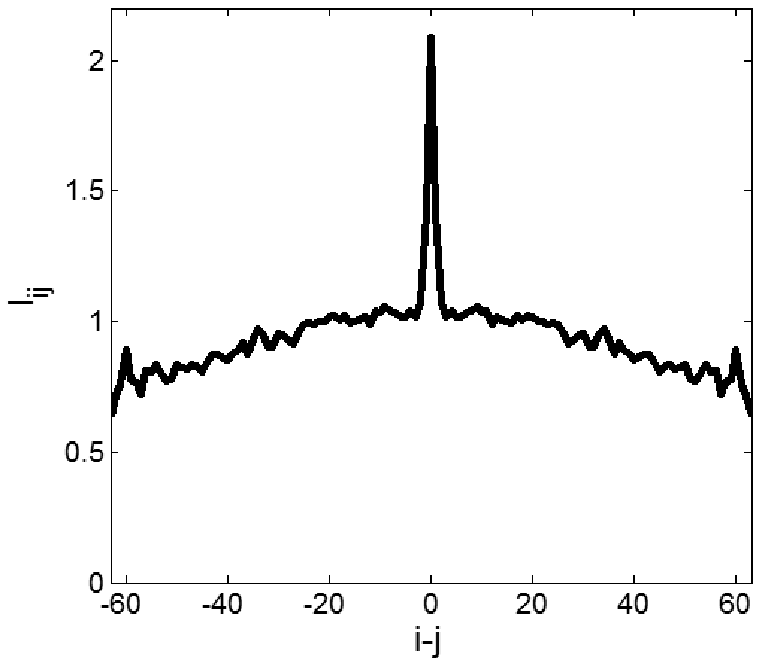}}}%
\caption{\label{fig:fig4}(a) Mean intensity $I_{ij}$ of coefficients $\tilde{k}_{ij}$, normalized by $\frac{1}{N}$. Results have been averaged over all times $T$ and frequencies $f$. (b) $I_{ij}$ as a function of $i-j$ (an average is achieved over couples $(i,j)$ such that $(i-j)=$constant).}
\end{minipage}
\end{center}
\end{figure}
Actually, Fig.\ref{fig:fig4}(a) shows that $ \sigma_{ij}^2$ is not uniform over the pairs $(i,j)$. The intensity $I_{ij}=\left < \left |\tilde{k}_{ij} \right |^2 \right >_{T,f}$ averaged over time $T$ and frequency $f$ is displayed for each source/receiver couple $(i,j)$. It serves as an estimate of the variance $ \sigma_{ij}^2$. The diagonal elements $\tilde{k}_{ii}$ exhibit a doubled variance compared to the off-diagonal elements. Moreover, the variance of off-diagonal elements tends to decrease when the distance $\left | i-j \right |p$ between the source and the receiver increases. From these results, it appears that the variance of the matrix entries, instead of being the same for all pairs $(i,j)$, could be modelled as
\begin{equation}
\label{eqn:sigma_ij_1}
N\sigma_{ij}^2=N\left < \left |\tilde{k}_{ij} \right |^2 \right > \simeq \delta_{ij}+f(|i-j|)
\end{equation}
where $\delta$ is the Kronecker symbol and $f$ a slowly decaying function such that $f(0)=1$.

The doubled variance of diagonal elements compared to the off-diagonal ones is well explained by the \textit{coherent backscattering effect}\cite{Ishimaru1, akkermans, wolf}. It corresponds to an enhancement (by a factor of 2) in the intensity of waves scattered in the backward direction. This phenomenon, also known as weak localization, originates from a constructive interference between a wave traveling along a multiple-scattering path and its reciprocal counterpart: it appears when multiple scattering occurs, as long as the reciprocity symmetry is preserved. When source and receiver are identical $(i=j)$, any path $q$  whose first and last scatterers are of coordinates $\left (X_q^{(1)},Z_q^{(1)}\right )$ and $\left (X_q^{(2)},Z_q^{(2)}\right )$, interfere constructively with its reciprocal counterpart whose first and last scatterers are located respectively at $\left (X_q^{(2)},Z_q^{(2)}\right )$ and $\left (X_q^{(1)},Z_q^{(1)}\right )$. As a consequence, on average, the diagonal elements $\tilde{k}_{ii}$ have a variance twice that of off-diagonal elements. Actually, this effect could also affect the matrix elements close to the diagonal. The typical width of the coherent backscattering peak is $\frac{a}{k\sqrt{D(T-2a/c)}}$ \cite{tourin,tourin2}, which has to be compared to the array pitch. In this experimental configuration, we have $T>70$ $\mu s$, $D \simeq 4 mm^2/\mu s$ \cite{aubry}, $a= 25$ $mm$ hence  $\frac{a}{k\sqrt{D(T-2a/c)}} < 0.16$ mm, whereas the array pitch is $p=0.417$ mm. Therefore the enhancement due to coherent backscattering is strictly limited to the diagonal elements. Finally, upon renormalization (Eq. \ref{eqn:eq2}), taking the coherent backscattering effect into account amounts to rewrite the variance of $\tilde{k}_{ij}$ as $\sigma_{ij}^2=\frac{1+\delta_{ij}}{N+1}$ (see Appendix \ref{app:var_double}).

Now, one can wonder if the doubled variance of diagonal elements $\tilde{k}_{ii}$ has any influence on the distribution of the singular values.
It is shown in Appendix \ref{app:var_double} that if the matrix dimension $N$ is large enough, the influence of the coherent backscattering peak can be neglected as long as it is limited to the diagonal elements $\left(\frac{a}{k\sqrt{DT}}<p\right)$, provided that the matrix is renormalized according to Eq.\ref{eqn:eq2}. We will therefore consider that the coherent backscattering enhancement has no significant effect on the distribution of singular values. 

The doubled variance along the diagonal of $\mathbf{\tilde{K}}$ is not the only deviation from the identical distribution assumption. Indeed, if we focus on the dependence of $\sigma_{ij}^2$ as a function of $(i-j)$ (see Fig.\ref{fig:fig4}(b)), we observe that it is not uniform: it decreases with the distance $x=|i-j|p$ between the source $i$ and the receiver $j$. This is due to the progressive growth of the diffusive halo (i.e the mean intensity) inside the medium. It is not instantaneous but depends on the diffusion contant $D$\cite{ishimaru}. In a near-field configuration($a<<Np$, $Np$ being the array size), $\sigma_{ij}^2$ would decrease as $\exp \left ( - \frac { x ^2 }{4DT} \right)$. And as soon as $\sqrt{DT}$ becomes significantly larger than the array size ($Np$), $\sigma_{ij}^2$ can be considered as constant \cite{mamou}. In a far-field situation ($a>>Np$), $\sigma_{ij}^2$ is always constant, as long as $i \neq j$. The experimental case we study corresponds to an intermediate configuration between near and far field. 

\begin{figure}[htbp] 
\begin{center}
\begin{minipage}{140mm}
\subfigure[]{
\resizebox*{7cm}{!}{\includegraphics{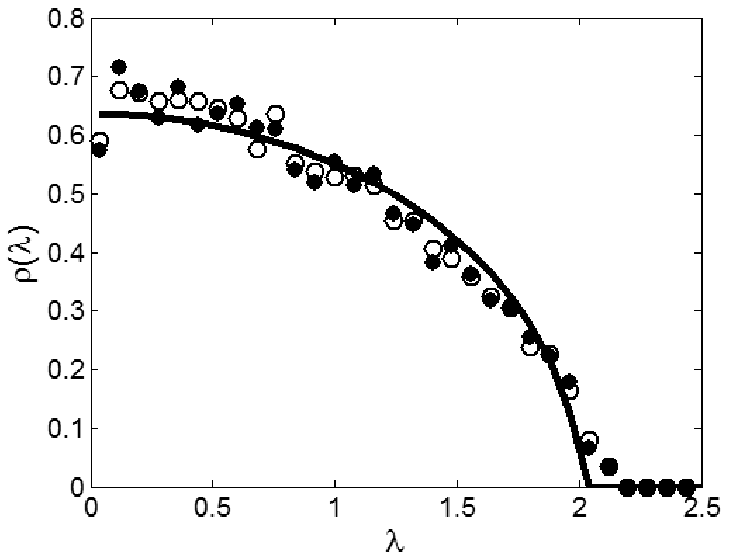}}}%
\subfigure[]{
\resizebox*{7cm}{!}{\includegraphics{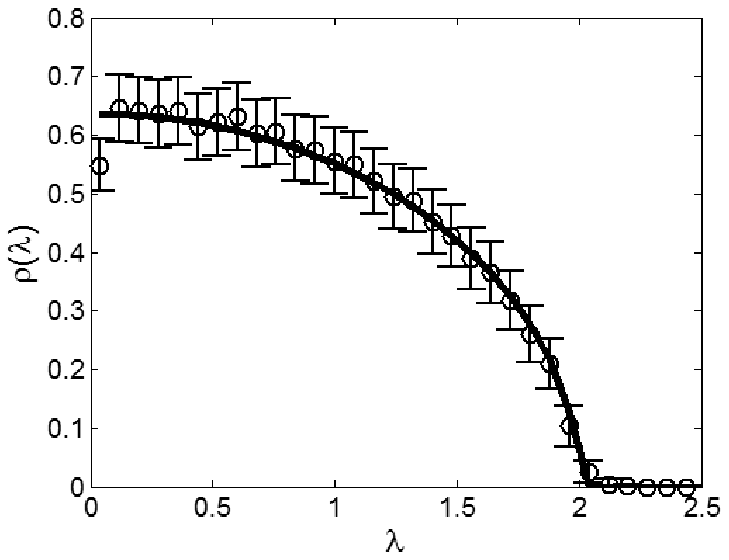}}}%
\caption{\label{fig:fig5}(a)Influence of the non-uniform variance of $k_{ij}$: The experimental distributions of singular values for $\mathbf{\tilde{K}_t}$(black disks) and $\mathbf{\tilde{K}^C_t}$(white disks) are compared to the quarter circle law (black continuous line). (b) Influence of the symmetry of $\mathbf{K}$: the result of the numerical simulation (white disks) is compared to the quarter-circle law (black continuous line). The error bars are $\pm$ two standard deviations.}
\end{minipage}
\end{center}
\end{figure}

A solution to deal with the variations of $\sigma_{ij}^2$ is to compensate for them. $\sigma_{ij}^2$ depends only on time $T$ and $|i-j|$ (Eq.\ref{eqn:sigma_ij_1}). Thus we can estimate $\sigma_{ij}^2$ by averaging $\left | \tilde{k}_{ij}(T,f) \right |^2$ over frequency $f$ and elements $(i,j)$ which are separated by the same amount $m=|i-j|$ :
\begin{equation}
\widehat{\sigma^2}(T,m)=\left < \left | \tilde{k}_{ij}(T,f) \right |^2 \right >_{f,\{(i,j)\,|\, m=|i-j|\}}
\end{equation}
where $\widehat{\sigma^2}(T,m)$ is the estimator of $\sigma_{ij}^2$. Then, at each time $T$, matrix entries $\tilde{k}_{ij}$ are normalized once again as:
\begin{equation}
\tilde{k}^C_{ij}(T,f)=\frac{\tilde{k}_{ij}(T,f)}{\sqrt{\widehat{\sigma^2}(T,|i-j|)}}
\end{equation}
A set of compensated matrices $\mathbf{\tilde{K}^C}(T,f)$ is built from $\mathbf{\tilde{K}}$. One can show that these matrices satisfy the identical distribution property : the variance of elements $\tilde{k}^C_{ij}$ is now constant over all pairs $(i,j)$. Once this operation is achieved, the new distribution of singular values can be investigated and is plotted in Fig.\ref{fig:fig5}(a). However, the two sets of matrices $\mathbf{\tilde{K}_t}$ and $\mathbf{\tilde{K}_t^C}$ lead to similar singular value spectra. It means that in our case the variations of $\sigma_{ij}^2$ with $|i-j|$ are not sufficiently large to significantly modify the distribution of the singular values. Nevertheless, in other experimental configurations, the non-uniformity of $\sigma_{ij}^2$ could have a stronger influence. For instance, in a near-field configuration, the growth of the diffusive halo is more visible and the distribution of singular values would exhibit stronger deviations from the quarter circle law, specially at early times.

\subsection{\label{subsec:correlation}Influence of correlations on the singular spectrum of $\mathbf{\tilde{K}}$}
As we can see in Fig.\ref{fig:fig3}(a), the experimental matrix $\mathbf{\tilde{K}}(T,f)$ exhibits correlations between adjacent entries. There are two reasons for that. First, there is a mechanical coupling between neighboring array elements. Second, the wave recorded on the array can be seen as the radiation of a spatially incoherent source with width $W$ (the diffuse halo inside the multiple-scattering medium) observed at a distance $a$. The Van Cittert-Zernike theorem\cite{goodman,derode3} states that the typical coherence length of the wave field is $\lambda a/W$ (in other words, the waves radiated by a finite-size incoherent source see their coherence length increases as they propagate). These two effects result in a short-range correlation between the scattered signals recorded on the array. In the experimental situation we investigated, the residual correlations are limited in range to adjacent elements, both in emission and reception, as it is shown in Fig\ref{fig:fig3}(a). They are mainly due here to the mechanical coupling between neighboring array elements, the coherence length of the diffuse wave-field becoming rapidly smaller than the array pitch in our experimental configuration.

Sengupta and Mitra\cite{sengupta} have investigated theoretically the influence of correlations on the distribution of singular values in the case of large-dimension random matrices. Their model assumes first that $\left < \tilde{k}_{ij} \right >=0$. The correlation between two coefficients $\tilde{k}_{il}$ and $\tilde{k}_{jm}$ is expressed as 
\begin{equation}
\label{eqn:eq6}
\left < \tilde{k}_{il}\tilde{k}_{jm}^*\right>=N^{-1}c_{ij}d_{lm}
\end{equation}
where the symbol $<.>$ denotes an ensemble average. $\mathbf{C}$ and $\mathbf{D}$ are $N \times N$ matrices. We will refer to them as the correlation matrices. Based on Eq.\ref{eqn:eq6}, Sengupta and Mitra predict the distribution of singular values using diagrammatic and saddle point integration techniques. Only the eigenvalues of $\mathbf{C}$ and $\mathbf{D}$ are required to obtain the singular values distribution of $\mathbf{K}$.

We now apply the Sengupta and Mitra's approach to our experimental configuration. Here, the coefficients $c_{ij}$ and $d_{ij}$ are identical because of spatial reciprocity : $\mathbf{C} \equiv \mathbf{D}$. We estimate $c_{ij}$ by the correlation coefficient $\Gamma_{i-j}$ defined in Eq.\ref{eqn:corr_coeff} :
\begin{equation}
\label{eqn:eq7}
\hat{c}_{ij}=\Gamma_{i-j}
\end{equation}
The coefficients $\hat{c}_{ij}$ form a matrix $\mathbf{\hat{C}}$ which is an estimator of the correlation matrix $\mathbf{C}$. $\mathbf{\hat{C}}$ is a Toeplitz matrix : its coefficients only depend on $i-j$. The application of Sengupta and Mitra's approach to our experimental situation embodies two assumptions :
\begin{itemize}
\item The correlation between two signals received(emitted) by the same element $l$ and emitted(received) by two elements $i$ and $j$ does not depend on the element $l$ but only on the distance $|i-j|$.
\item The correlation between two signals emitted and received by two pairs of elements $(i,l)$ and $(j,m)$ is given by the product of correlations at emission $\left (\Gamma_{i-j}\right)$ and at reception $\left (\Gamma_{l-m}\right)$.
\end{itemize}
The first assumption requires the random medium to be statistically invariant by translation. The second one is made for analytical tractability and is verified in our case.

Once the correlation coefficient $\Gamma_m$ is measured experimentally at emission/reception (see Fig.\ref{fig:fig3}(a)), the correlation matrix $\mathbf{C}$ is estimated by $\mathbf{\hat{C}}$ (Eq.\ref{eqn:eq7}). Eigenvalues of $\mathbf{\hat{C}}$ are calculated numerically and incorporated in Sengupta and Mitra's model in order to obtain an estimation of the singular value distribution for $\mathbf{\tilde{K}}$. 

In Fig.\ref{fig:fig3}(b), the theoretical result provided by Sengupta and Mitra's method is compared to the experimental singular value distribution but also to the result provided by a numerical simulation. It consists in generating numerically a matrix $\mathbf{P}$ whose elements are circularly symmetric complex gaussian random variables with zero mean. Then, a matrix $\mathbf{Q}$ is built from $\mathbf{P}$, such that
\begin{equation}
\label{eqn:eq8}
\mathbf{Q}=\mathbf{\hat{C}}^{\frac{1}{2}}\mathbf{P}\mathbf{\hat{C}}^{\frac{1}{2}}
\end{equation}
One can show that the matrix $\mathbf{Q}$ exhibits the same correlation properties at emission and reception as the experimental matrix $\mathbf{K}$. A histogram of singular values is obtained by achieving the SVD of $\mathbf{Q}$, renormalizing its singular values according to Eq.\ref{eqn:eq2} and averaging the result over 2000 realizations. The agreement between the numerical and theoretical distributions of singular values is perfect, which illustrates the validity of the theoretical method given in \cite{sengupta}. Taking the correlations into account improves significantly the agreement between theory and experiment. 

{As was mentioned earlier, when analysing the experimental data we crudely suppressed the correlations between neighbours by removing from the matrix most of its elements. One can object that it is silly to throw away the data, instead of trying to make the most of it. It is true that the correlations could have been cancelled out by applying the following operation
\begin{equation}
\label{eqn:decorrelation}
\mathbf{R}=\mathbf{\hat{C}^{-1/2} \tilde{K} \hat{C}^{-1/2}}.
\end{equation}
$\mathbf{R}$ has uncorrelated elements, while keeping the same size as $\mathbf{K}$. This procedure amounts to ``whitening'' the data. The singular value distribution of $\mathbf{R}$ would then fit the quarter-circle law. However, this method would be relevant only if the correlations were exclusively due to the experimental device (here, a mechanical coupling between neighbouring transducers). As mentioned previously, there also exists a physical correlation due to the intrinsic coherence of the diffuse wave-field. And in Sec.\ref{sec:ss} we will wee that there may be another form of correlation, a long-range correlation, that is typical of single scattering.
In the general case of an unknown medium, single and multiple scattering coexist, and all correlations (short- and long-range) must not be cancelled out. Indeed, applying Eq.\ref{eqn:decorrelation} blindly would affect equally single and multiple scattering contributions, whereas their statistical properties differ (see Sec.\ref{sec:ss}). This is particularly important
for target detection in a random medium (see Sec.\ref{sec:detect_target}), where one tries to detect the
single-scattered echo from a target drowned in multiple scattering, based on the statistics of the singular values.
}

In addition to short-range correlations between neighbouring elements, the matrix $\mathbf{K}$ is symmetric because of spatial reciprocity. This property results once again in a deviation from the quarter circle law theorem which assumes independent matrix coefficients \cite{tulino}. We will consider here the case of the truncated matrix $\mathbf{\tilde{K}}_t$ as defined in Sec.\ref{subsubsec:exp_obs_ms}, in order to study the effect of symmetry independently from the presence of correlations between adjacent entries. Another numerical simulation is performed. It consists in generating a symmetric matrix $\mathbf{P}$ of dimension $\frac{N}{2} \times \frac{N}{2}$ whose elements are circularly symmetric complex gaussian random variables with zero mean. A histogram of singular values is obtained by computing the SVD of $\mathbf{P}$, renormalizing its singular values according to Eq.\ref{eqn:eq2} and averaging the result over 2000 realizations. The distribution of singular values provided by numerical calculations does not exactly match the quarter circle law (see Fig.\ref{fig:fig5}(b)). More precisely, the deviation appears mainly in the vicinity of $\lambda=0$. The reason for this deviation is shown in Appendix \ref{app:sym}. By and large, the symmetry of $\mathbf{K}$ induces additional correlations between diagonal elements of the autocorrelation matrix $\mathbf{K}\mathbf{K}^{\dag}$, which do not exist when the random matrix is not symmetric. These residual correlations are shown to be of the order of $\frac{1}{N}$. In our experimental configuration ($N/2=32$), they are quite low but sufficient to induce a slight deviation from the quarter circle law especially near $\lambda=0$ which can also be observed in the experimental results (Fig.\ref{fig:fig5}a).

As to the other end of the singular value spectrum, an interesting theoretical result is that the distribution of singular values should remain bounded (for $N>>1$) even in presence of correlations. If correlations are ignored, the maximum value predicted by the quarter-circle law (which would apply if the matrix elements were independent and $N >> 1$) is $\lambda_{max}=  2$. Here, taking $\Gamma_m$ into account, theoretical results show that the singular values cannot exceed $\lambda_{max}=  2.5$ (see Fig.\ref{fig:fig3}(b)).
However, note that a slight difference between experimental and theoretical results remains near $\lambda_{max}$ (see Figs.\ref{fig:fig2}, \ref{fig:fig3}(b) \& \ref{fig:fig5}(a)). A possible origin for this small discrepancy is the presence of recurrent paths, \textit{i.e.}, multiple-scattering paths for which the first and last scatterers are identical or located in the same resolution cell. A partial wave associated with a recurrent path would display the same statistical behavior as single scattered waves, similarly to what happens in the coherent backscattering phenomenon, where recurrent paths tend to diminish the enhancement factor \cite{wiersma}. Therefore, as the single scattering contribution results in a singular value distribution which is not of  bounded support (see Sec.\ref{sec:ss}), recurrent scattering may be responsible for the slight disagreement between experimental and theoretical results near $\lambda_{max}$. However, the remaining disagreement between theory and experiment is small, and it is difficult to go beyond speculations about its origin.

\subsection{\label{subsec:ms_conclu}Conclusion}
It appears that in the absence of spatial correlations between matrix entries $k_{ij}$, the experimental distribution of singular values is close to the quarter circle law. Nevertheless, some deviations from the assumptions generally made in RMT have been pointed out. First, the elements of $\mathbf{K}$ are not identically distributed. The reason for that is the inhomogeneous distribution of backscattered intensity, due to the coherent backscattering enhancement as well as the decreasing of the diffuse halo with the distance between the transmitter and the receiver. Another cause of deviation has been pointed out numerically: the symmetry of the array response matrix, which induces additional correlations between matrix coefficients. All these effects become negligible as long as $N>>1$. Yet they slow down the convergence of $\rho(\lambda)$ to the quarter-circle law. 

In the presence of correlations between entries, the method proposed by Sengupta and Mitra\cite{sengupta} allows one to calculate the distribution of singular values. As a preliminary conclusion, in the case where multiple scattering dominates, the normalized response matrix $\mathbf{\tilde{K}}$ falls into the general scope of RMT and a theoretical prediction of the singular value spectrum can be achieved. Particularly, an upper bound $\lambda_{max}$ for the singular values can be calculated. As it will be shown in the next section, this is no longer true if single scattering dominates. Note also that this agreement with classical results of RMT only holds in a diffusive regime. Indeed, if we approach Anderson localization ($k l_e \sim 1$), one can expect that interference effects, such as hot spots \cite{Nieuwenhuizen,hu} or loops \cite{wiersma,haney}, should strongly affect the distribution of singular values. Yet in our experimental configuration, we are far from strong localization ($k l_e \sim 100$).

\section{\label{sec:ss}Single-scattering regime}
\begin{figure}[htbp] 
\begin{center}
\begin{minipage}{140mm}
\subfigure[]{
\resizebox*{7cm}{!}{\includegraphics{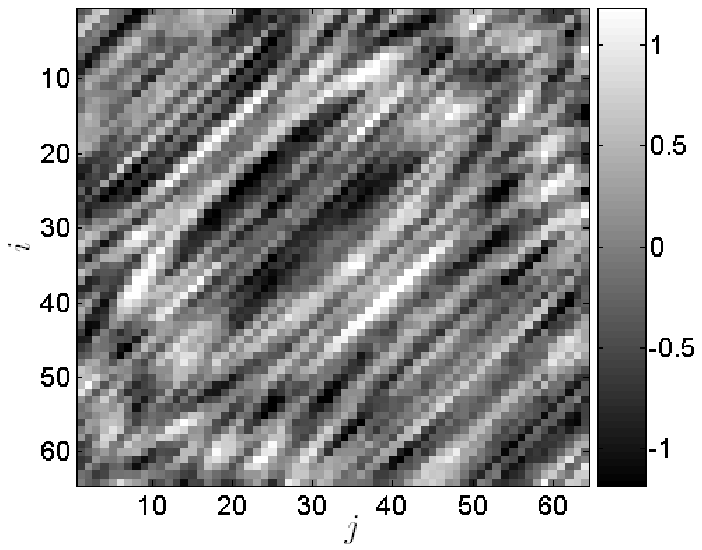}}}%
\subfigure[]{
\resizebox*{7cm}{!}{\includegraphics{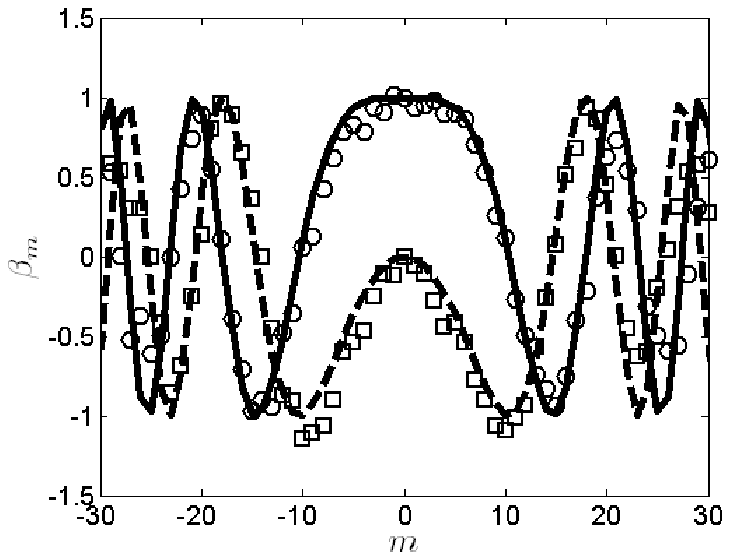}}}%
\caption{\label{fig:fig6}Experimental results in a single-scattering sample (agar gel). (a) Real part of the matrix $\mathbf{\tilde{K}}(T,f)$ obtained at time $T$=265 $\mu s$ and frequency $f$=3.1 MHz. (b) Real(white disks) and imaginary(white squares) parts of $\beta_{m}$ are shown as a function of $m$ and are compared to the real(continous line) and imaginary (dashed line) parts of Eq.\ref{eqn:eq_ss_signal_3}.}
\end{minipage}
\end{center}
\end{figure}

In this section, we consider the case of a weakly scattering medium. The sample under investigation is now a slab of gel (composed of 5\% of gelatine and 3\% of agar-agar), with thickness $L\simeq 100$ mm and a mean free path $l_e \sim 1000$ mm. In such conditions, the multiple scattering contribution is negligible. The array-sample distance $a$ is 60 mm. The experimental procedure and the numerical processing are performed as described in Sec.\ref{sec:exp_num}. A typical example of a matrix $\mathbf{\tilde{K}}(T,f)$ measured experimentally is given in Fig.\ref{fig:fig6}(a). Contrary to the multiple scattering regime, the array response matrix $\mathbf{\tilde{K}}$ exhibits a deterministic coherence along its antidiagonals, although the scatterers distribution in the agar gel is random.
We briefly recall the origins of this phenomenon \cite{aubry09} and its impact on the singular values distribution. 

As before, the signals $k_{ij}(T, f)$ at a time $T$ and frequency $f$ correspond to the sum of partial waves that reach the array in the time window $[T-\Delta t/2 ;T+\Delta t/2]$ except that only single-scattering paths are now considered. The \textit{isochronous volume} is defined as the ensemble of points that contribute to the backscattered signal at a given time. It is formed by a superposition of ellipses whose foci are transmitter $i$ and receiver $j$. In a far-field configuration, we approximate the isochronous volume by a slab of thickness $\Delta r=c\Delta t$, located at a distance $R=cT$ from the array and parallel to it. For simplicity but without loss of generality, we also assume that the reflectors as well as the array elements are point-like. 

In a 2D configuration, under the paraxial approximation, $k_{ij}(T,f)$ can be expressed as
\begin{equation}
\label{eqn:eq_ss_signal}
k_{ij}(T,f) \propto \frac{\exp \left ( j 2kR \right )}{R}\sum_{d=1}^{N_d} A_d \exp \left [ j k \frac{ \left (x_i -X_d \right )^2}{2R} \right ] \exp \left [ j k \frac{ \left (x_j -X_d \right )^2}{2R} \right ]
\end{equation}
The index $d$ denotes the $d^{th}$ path which contributes to the signal received at time $T$. $X_d$ is the transversal position of the reflector associated with this path and the amplitude $A_d$ depends on the reflectivity of the scatterer. 

Let us express $k_{ij}$ as a function of variables $(x_i-x_j)$ and $(x_i+x_j)$:
\begin{equation}
\label{eqn:eq_ss_signal_2}
k_{ij}(T,f) \propto  \underbrace { \frac{\exp \left ( j 2kR \right )}{R} \exp \left [ j k \frac{ \left (x_i -x_j \right )^2}{4R} \right ] }_{\mbox{deterministic term}}  \underbrace{ \sum_{d=1}^{N_d} A_d  \exp \left [ j k \frac{ \left (x_i + x_j - 2X_d \right )^2}{4R} \right ] }_ {\mbox{random term}} 
\end{equation}
The term before the sum in Eq.\ref{eqn:eq_ss_signal_2} does not depend on the scatterers distribution. 
On the contrary, the term on the right does; hence it is random. As a consequence,  single scattering manifests itself as a particular coherence along the antidiagonals of the matrix $\mathbf{K}$, as illustrated in Fig.\ref{fig:fig6}. Indeed, in a given sample, along each antidiagonal (\textit{i.e.} for couples of transmitter $i$ and receiver $j$ such that $i + j$ is constant), the random term of Eq.\ref{eqn:eq_ss_signal_2} is the same. So, whatever the realization of disorder, there is a deterministic phase relation between coefficients of $\mathbf{K}$ located on the same antidiagonal. It can be written as:
\begin{equation}
\label{eqn:eq_ss_signal_3}
\beta_{m}=\frac{k_{i-m,j+m}(T,f)}{k_{ii}(T,f)} =  \exp \left [ j k \frac{ \left (m p \right )^2}{R} \right ]
\end{equation}
Note that this essential result is valid independently of the scatterers configuration, under two conditions: single scattering and paraxial approximation. The parabolic phase dependence along each antidiagonal predicted by Eq.\ref{eqn:eq_ss_signal_3} is compared in Fig.\ref{fig:fig6}(b) with the coefficient $\beta_{m}$ obtained experimentally at time $T=265$ $\mu s$ and frequency $f=3.1$ MHz along the main antidiagonal (\textit{i.e} $i=32$). Theoretical and experimental results are in a very good agreement.

\begin{figure}[htbp] 
\begin{center}
\includegraphics{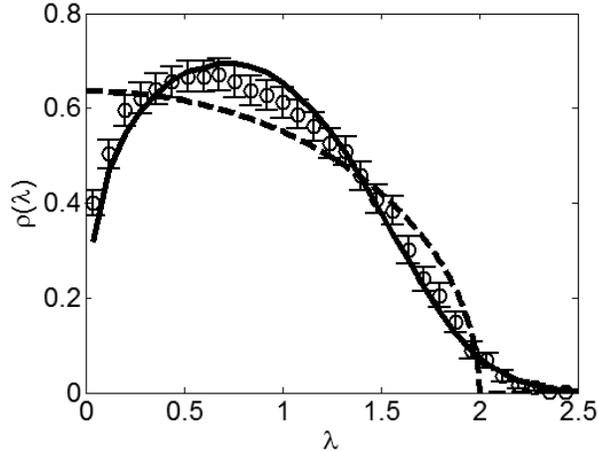}
\caption{\label{fig:fig7}The experimental distribution of singular values $\hat{\rho}(\lambda)$ in the single-scattering case (white disks) is compared to the quarter circle law $\rho_{QC}(\lambda)$ (dashed curve) and to the the Hankel law $\rho_H(\lambda)$(continuous curve) which has been calculated numerically. The error bars are $\pm$ two standard deviations.}
\end{center}
\end{figure}

In order to investigate independently the effect of the deterministic coherence along antidiagonals of $\mathbf{\tilde{K}}$, other correlations that may exist between lines and columns of the matrix $\mathbf{K}$ have to be removed. These correlations are measured by assessing the normalized correlation coefficient $\Gamma_m$ (Eq.\ref{eqn:corr_coeff}). $\Gamma_m$ (not shown here) spreads until $|m|=2$ for this experiment. Thus, the initial set of matrices $\mathbf{K}(T,f)$ has been truncated by keeping only one in three elements. Note that only correlations between lines and columns of $\mathbf{K}$ are cleared. The deterministic coherence along antidiagonals is long-range and hence is not removed by this operation. As before, the truncated matrix $\mathbf{K_t}(T,f)$ is renormalized (Eq.\ref{eqn:eq2}). The experimental distribution of singular values is shown in Fig.\ref{fig:fig7}: clearly, it does not follow the quarter circle law, even though correlations between neighbouring entries have been removed.
This is due to the deterministic coherence of single scattered signals along the antidiagonals of $\mathbf{\tilde{K}}_t$. 

To our knowledge, this kind of random matrix whose antidiagonal elements are linked with a deterministic phase relation has not been yet investigated theoretically. But its properties are close to those of a random Hankel matrix, whose spectral behaviour has been studied recently\cite{bryc}. A Hankel matrix is a $N \times N$ square matrix whose elements belonging to the same antidiagonal ($i+j=$ constant) are equal. Let $\left \{ a_p \right \}$ be a sequence of $2N-1$ complex random variables identically and independently distributed with zero mean. The Hankel matrix $\mathbf{R}$ built such as $r_{ij}=a_{i+j-1}$, is said random. We have checked numerically that a random matrix whose antidiagonal elements are linked with a deterministic phase relation displays the same singular value distribution as a Hankel random matrix, provided that all elements of this matrix are zero mean and have the same variance. In the literature, Bryc \textit{et al.} \cite{bryc} have proved, for normalized random Hankel matrices, the almost sure weak convergence of the distribution of eigenvalues to a universal, non random, symmetric distribution of unbounded support. We will assume that this convergence property also applies to the singular values distribution of a random Hankel matrix.
In the following, $\rho_H(\lambda)$ will denote the asymptotic singular values distribution of a random Hankel matrix and will be referred to as the Hankel law. To our knowledge, no analytical expression of the Hankel law has been found yet and only a numerical calculation can provide an estimate of $\rho_H(\lambda)$. In Fig.\ref{fig:fig7}, the experimental distribution of singular values of $\mathbf{\tilde{K}}_t$ has been compared to the Hankel law. $\rho_H(\lambda)$ is provided by a numerical generation of random Hankel matrices. The agreement between both curves is satisfactory. An important feature of the Hankel law is its unbounded support. As a consequence, in the single scattering regime, the first singular value has no bound, contrary to the multiple scattering case for which the first singular value could never be higher than $\lambda_{max}$ in the asymptotic limit ($N \rightarrow \infty$).

\section{\label{sec:detect_target}Probability distribution of $\tilde{\lambda}_1$}
So far, we have studied the distribution of singular values as a whole, in relation with the importance of single or multiple scattering within the medium. We now focus on the strongest singular value, $\tilde{\lambda}_1$. Knowing its probability distribution function is crucial for applications to  detection. Imagine a random scattering medium, in which a target (\textit{i.e.}, a stronger
reflector) may be embedded. Once $\mathbf{K}$ is measured, if the strongest normalized singular value $\tilde{\lambda}_1$ is above a certain threshold $\alpha$, we will conclude (with a certain probability of error) that there is indeed a target. In order to assess the performance of a detection scheme based on a SVD of the propagation matrix, we need a reliable model for the pdf of $\tilde{\lambda}_1$.
%

Assume the occurrence of a target at a depth $R$. The array response matrix $\mathbf{K}(T,f)$ at the time of
flight $T = 2R/c$ can be written as :
\begin{equation}
\label{eqn:decompos_K}
\mathbf{K}(T,f)=\mathbf{K^T}(T,f)+\mathbf{K^R}(T,f)
\end{equation}
$\mathbf{K^T}(T,f)$ is the contribution of the direct echo reflected by the target, and $\mathbf{K^R}(T,f)$ the response of the random medium, which may include single scattering, multiple scattering contributions as well as additive noise. $\mathbf{K^R}(T,f)$ can be seen as a perturbation (not necessarily small) of $\mathbf{K^T}(T,f)$. As usual, the matrix is renormalized according to Eq.\ref{eqn:eq2}. 
For simplicity, we will assume that short-range correlations have been removed, so that the relevant probability density function of the singular values of $\mathbf{K^R}$ should be the quarter-circle law (for $N>>1$ in the multiple scattering regime) and the Hankel law (in the single scattering regime).

In a first approximation, $\mathbf{K^T}$ is of rank 1 (actually, for a resonant or a large target $\mathbf{K^T}$ may have more than one significant singular value \cite{chambers,minonzio2,zhao,aubry06,robert}). $\mathbf{K^R}$ is random, and its normalized singular values have a probability density $\rho^R(\lambda)$. Depending on the scattering properties of the medium (particularly, its mean-free path), at time $T=2R/c$, $\rho^R(\lambda)$ may follow the Hankel law (if single scattering dominates), the quarter-circle law (if multiple scattering or additive noise dominate) or a combination of both in intermediate situations. Let $\tilde{\lambda}_1^R$ denote the highest singular value of $\mathbf{\tilde{K}^R}(T,f)$. As a detection threshold based on the first singular value of $\mathbf{K}$ is needed, the statistical behavior of $\tilde{\lambda}_1^R$ has to be known. If the singular values of $\mathbf{K^R}(T,f)$ were independent from each other, then the distribution function $F_1^R$ of $\tilde{\lambda}_1^R$ would be simply given by the $N^{th}$ power of the distribution function $F^R(\lambda)$ of one singular value, with
$$F^R(\lambda)= \int_0^{\lambda}dx\rho^R(x)\mbox{.}$$
Actually, the singular values of a random matrix are not independent because of level repulsion \cite{mehta,pastur}. It implies a zero probability for degenerate singular spaces: the singular values tend to keep away from each other. Thus, the probability density function $\rho_1^R(\lambda)$ and the distribution function $F_1^R(\lambda)$ of the first singular value $\tilde{\lambda}_1^R$ cannot be deduced simply from $\rho^R(\lambda)$.

\begin{figure}[htbp] 
\begin{center}
\begin{minipage}{140mm}
\subfigure[]{
\resizebox*{7cm}{!}{\includegraphics{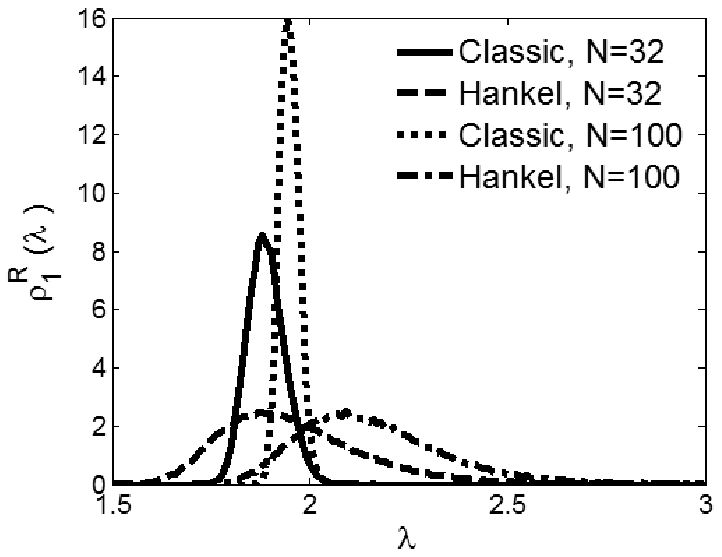}}}%
\subfigure[]{
\resizebox*{7cm}{!}{\includegraphics{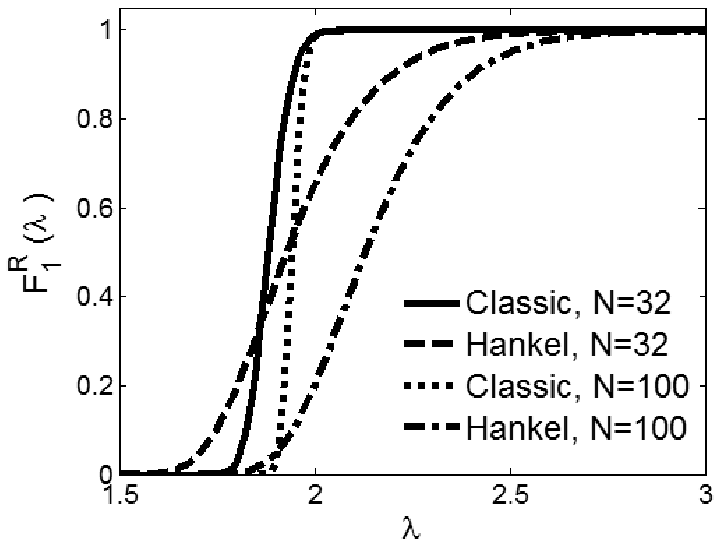}}}%
\caption{\label{fig:fig8} (a) Probability density function $\rho_1^R$ of the first singular value $\tilde{\lambda}^R_1$ estimated numerically. The multiple scattering case corresponds to continuous ($N=32$) and dotted ($N=100$) curves. The single-scattering case corresponds to dashed ($N=32$) and dash-dotted ($N=100$) curves. (b) Distribution function $F_1^R(\lambda)$ of the first singular value $\tilde{\lambda}^R_1$. The multiple-scattering case  still corresponds to continuous ($N=32$) and dotted ($N=100$) curves. The single-scattering case corresponds to dashed ($N=32$) and dash-dotted ($N=100$) curves.}
\end{minipage}
\end{center}
\end{figure}

Once again, we will refer to RMT and particularly to the theoretical studies dealing with the first eigenvalue of the autocorrelation matrix $\mathbf{K^RK^{R\dag}}$, \textit{i.e} $\left [ \tilde{\lambda}_1^R \right ]^2$. If $\mathbf{K^R}$ is a ``classical'' random matrix (multiple scattering regime), $\left [ \tilde{\lambda}_1^R \right ]^2$ is given by \cite{tracy2,tracy,johansson,johnstone,karoui2}:
\begin{equation}
\label{eqn:rho1_RMT}
\left [ \tilde{\lambda}_1^R \right ]^2 = 4 + \left ( \frac{N}{4} \right)^{- \frac{2}{3}} Z+ o\left ( N^{-\frac{2}{3}}\right)
\end{equation}
where $Z$ is a random variable whose probability density function is a complicated law, known as the \textit{Tracy Widom} distribution. It is an asymmetric bell curve, uncentered and of infinite support \cite{tracy2,tracy}. No analytic expression is available, nevertheless it can be estimated numerically. In our case, we are rather interested in the probability density function $\rho_1^{R}$ of the first singular value $ \tilde{\lambda}_1^R $. $\rho_1^{R}$ has been estimated by generating numerically ``classical'' random matrices and is displayed in Fig.\ref{fig:fig8}(a), for $N=32$ and $N=100$. Although the quarter circle law $\rho_{QC}$ is of bounded support, $\rho_1^{R}$ is of infinite support. Indeed, $\rho_{QC}$ is only an asymptotic law, \textit{i.e} valid for $N \rightarrow \infty$. For a random matrix of finite dimension, $\tilde{\lambda}_1^R$ has a non zero probability of exceeding $\lambda_{\mbox{\small max}}=2$. Nevertheless, $\rho_1^{R}(\lambda)$ narrows with $N$: the variance of $\tilde{\lambda}_1^R$ decreases when $N$ grows, and its expected value has been shown to be bounded by $\lambda_{\mbox{\small max}}=2$.

The relevant quantity for the detection issue is the distribution function $F_1^R$ of the first singular value $\tilde{\lambda}_1^R$. $F_1^R$ is the primitive of $\rho_1^R$,
$$F_1^{R}(\lambda)= P\left \{ \tilde{\lambda}_1^R \leq \lambda \right \}=\int_0^{\lambda}dx\rho^{R}_1(x)\mbox{.}$$
In terms of target detection, the probability of false alarm is $PFA(\alpha) =1-F_1^R(\alpha)$. By setting an acceptable $PFA$, we can determine the corresponding threshold above which a target is said to be detected. The distribution functions $F^{R}_1$, for $N=32$ and $N=100$, are plotted in Fig.\ref{fig:fig8}(b).
In the multiple scattering regime (``classical'' random matrix), we observe that, for $N=32$ and $N=100$, $F^{R}_1 \left ( \lambda =2 \right ) \simeq 0,99$: if the PFA is fixed at 1\%, $\alpha=2$ is the corresponding threshold.

This is no longer the case if the propagation operator behaves as a Hankel matrix (single scattering regime): the threshold is much higher, making the detection more difficult, for the same $PFA$. Theoretical results indicate that the expected value of the first singular value grows as $\sqrt{\log(N)}$ \cite{meckes}. This is illustrated in Fig.\ref{fig:fig8}: $\rho_1^R$ and $F_1^R$ have been calculated by generating numerically random Hankel matrices for $N=32$ and $N=100$. As $N$ is enlarged the curves are shifted towards right, which implies a larger probability of false alarm for a given detection threshold $\alpha$. Other theoretical studies have shown that the fluctuations of $\tilde{\lambda}_1^R$ are inferior to $\sqrt{\log N}$ \cite{adamczak} and that, in the gaussian case, the variance of $\tilde{\lambda}_1^R$ remains bounded \cite{chatterjee}.

In a real situation, single and multiple scattering coexist; what is the relevant distribution function $F_1^R$? If multiple scattering dominates (\textit{i.e.}, the target depth is larger than a few mean-free paths) or if the perturbation $\mathbf{K^R}$ is additive uncorrelated noise, the probability of false alarm $PFA(\alpha)=1-F_1^R(\alpha)$ is deduced considering the distribution function $F_1^R$ obtained for a ``classical'' random matrix. If no \textit{a priori} information is available regarding the scattering medium, we
have to calculate $F_1^R$ for a ``classical'' random matrix and for a random Hankel matrix, deduce two values for the probability of false alarm $1-F_1^R(\alpha)$, and keep the highest.

In the view of  applications (\textit{e.g.}, non-destructive evaluation, target detection \textit{etc}.) an acceptable probability of false alarm $P_0$ is set first, hence the detection threshold : 
\begin{equation}
\label{eqn:threshold}
\alpha=F^{-1}_1(1-P_0)
\end{equation}
Once the usual normalization is performed, if $\tilde{\lambda}_{1}>\alpha$ a target is detected at time $T$ and frequency $f$. Conversely, if $\tilde{\lambda}_{1}<\alpha$ one cannot conclude about the possible presence of a target. The application of this detection criterion has been performed experimentally in a recent study \cite{aubry09_2}.

The detection criterion (Eq.\ref{eqn:threshold}) can be used to estimate the performance of the DORT method in a random medium (or in the presence of noise). Let $\sigma_T^2$ and $\sigma_R^2$ be the power of signals associated with the target and the random contribution. If the first singular value $\lambda_1$ of $\mathbf{K}$ is associated to the target echo, then the expected value of $\lambda_1$ is $\lambda_1^T$. In Appendix \ref{app:sing_value}, it is shown that $\lambda_1^T = N\sigma_T$, so
$$\mbox{E} \left \{ \lambda_1 \right \}=N \sigma_T \,\mbox{.}$$
We also show in Appendix \ref{app:sing_value} that the quadratic mean of the singular values can be expressed as
$$ \sqrt { \frac{1}{N}\sum_{p=1} \lambda_p^2} = \sqrt{N \left (\sigma_T^2 + \sigma_R^2 \right) }\, \mbox{.}$$ 
Upon normalization, we have 
$$\mbox{E} \left \{ \tilde{\lambda}_1 \right \} = \sqrt{N\frac{\sigma_T^2}{\sigma_T^2+\sigma_R^2}}\simeq \frac{\sigma_T}{\sigma_R}\sqrt{N} \,\,\,\,\mbox{,     
for}\,\,\,\, \sigma_T^2<<\sigma_R^2$$
and 
$\mbox{var} \left \{ \tilde{\lambda}_1 \right \}\simeq\frac{1}{2}$.
Even though the echo of the target may be very weak compared to the scattering contribution, DORT can be expected to detect it if $\tilde{\lambda}_1 > \alpha$, \textit{i.e.},
\begin{equation}
\label{eqn:detection}
\frac{\sigma_T}{\sigma_R} > \frac{F_1^{-1}\left ( 1-P_0 \right )}{\sqrt{N}}\mbox{.}
\end{equation}
The most favourable situation is when the quarter-circle law is valid (multiple scattering regime or additive white noise, and $N>>1$) : in that case $\alpha$ can be fixed to 2, which corresponds for $N=32$ to a probability of false alarm $P_0 \simeq 1 \%$, and typically the target will be detected if 
\begin{equation}
\label{eqn:detection_RMT_2}
\frac{\sigma_T}{\sigma_R}>\frac{2}{\sqrt{N}}\mbox{.}
\end{equation}
In other words, the weakness of the target may be compensated by an increasing number of array elements: the performance of the DORT method is improved as the square-root of the number of independent channels. However in the single scattering regime, the expected value of $\tilde{\lambda}_{1}^R$ grows as $\sqrt{\log N}$ \cite{meckes}. For a given $P_0$, we can assume that $\alpha$ increases also as $\sqrt{\log N}$. Hence, the performance of the D.O.R.T method grows as $\sqrt{\frac{N}{\log N}}$, which is significantly slower than $\sqrt{N}$.

\section{Conclusion}
In this article, the distribution of singular values of the array response matrix $\mathbf{K}$ in a random medium has been investigated, in a backscattering configuration. Once a judicious renormalization is achieved, the distribution $\rho(\lambda)$ obtained experimentally with ultrasonic waves in scattering samples is in very good agreement with theoretical predictions. Interestingly, $\rho(\lambda)$ is shown to differ significantly in the single and multiple scattering regimes. When multiple scattering dominates, as long as spatial correlations between matrix entries are negligible, $\rho(\lambda)$ is found to approach the quarter circle law. Correlations between matrix entries can also be taken into account: in that case, $\rho(\lambda)$ can be calculated following the method proposed by Sengupta and Mitra\cite{sengupta}. On the contrary, single scattering contributions exhibit a different behaviour: whatever the realisation of disorder, a deterministic coherence persists along each antidiagonal of the matrix $\mathbf{K}$. As a consequence, $\rho(\lambda)$ no longer follows the quarter-circle law and the singular spectrum of $\mathbf{K}$ becomes analogous to that of a Hankel matrix. These results have been applied to the detection of a target embedded in a random scattering media. Once the matrix is renormalized, knowing the distribution $\rho(\lambda)$ allows one to define a rigorous detection criterion based on the strongest singular value, which is expected to be associated to the target. The perspectives of this work are many. The results could be applied to all fields of wave physics where coherent transmit/receive arrays are available for imaging and detection (\textit{e.g.}, non destructive testing of scattering materials, underwater acoustics, landmine detection, seismology, radar/sonar, \textit{etc.})

\section{Acknowledgments}
The authors wish to acknowledge Josselin Garnier, Arnaud Tourin, Claire Prada, Julien de Rosny and Mathias Fink for fruitful discussions. They also wish to acknowledge the groupe de recherches IMCODE of CNRS (GDR 2253), Patricia Daenens for her technical help and Victor Mamou who made the steel rods samples.

\bibliographystyle{tWRM}

\begin{thebibliography}{64}
\providecommand{\natexlab}[1]{#1}

\bibitem[1]{tulino}
A. Tulino and S. Verd\`{u}, {\itshape Random Matrix Theory and Wireless
  Communications}, Fundations and Trends in Communications and Information
  Theory 1 (2004), pp. 1--182.

\bibitem[2]{sprik}
R. Sprik, A. Tourin, J. {de Rosny}, and M. Fink, {\itshape Eigenvalue
  distributions of correlated multichannel transfer matrices in strongly
  scattering systems}, Phys. Rev. B 78 (2008), p. 012202.

\bibitem[3]{nuah}
C-N. Nuah, D. Tse, J. Kahn, and R. Valenzuela, {\itshape Capacity scaling in
  MIMO wireless systems under correlated fading}, IEEE Trans. Inform. Theory 48
  (2002), pp. 637--650.

\bibitem[4]{moustakas2}
A. Moustakas, S. Simon, and A. Sengupta, {\itshape MIMO capacity through
  correlated channels in the presence of correlated interferers and noise: A
  (not so) large N analysis}, IEEE Trans. Inform. Theory 49 (2003), pp.
  2545--2561.

\bibitem[5]{vellekoop}
I.M. Vellekoop and A.P. Mosk, {\itshape Universal optimal transmission of light
  through disordered materials}, Phys. Rev. Lett. 101 (2008), p. 120601.

\bibitem[6]{pendry}
J.B. Pendry, A.M. Kinnon, and P.J. Roberts, {\itshape Universality classes and
  fluctuations in disordered systems}, Proc. Roy. Soc. A 437 (1992), pp.
  67--83.

\bibitem[7]{prada}
C. Prada and M. Fink, {\itshape Eigenmodes of the time-reversal operator: a
  solution to selective focusing in multiple-target media}, Wave Motion 20
  (1994), pp. 151--163.

\bibitem[8]{prada2}
C. Prada, S. Manneville, D. Poliansky, and M. Fink, {\itshape Decomposition of
  the time reversal operator: Application to detection and selective focusing
  on two scatterers.}, J. Acoust. Soc. Am. 99 (1996), pp. 2067--2076.

\bibitem[9]{prada4}
C. Prada and J-L. Thomas, {\itshape Experimental subwavelength localization of
  scatterers by decomposition of the time reversal operator interpreted as a
  covariance matrix}, J. Acoust. Soc. Am. 114 (2003), pp. 235--243.

\bibitem[10]{minonzio}
J-G. Minonzio, C. Prada, A. Aubry, and M. Fink, {\itshape Multiple scattering
  between two elastic cylinders and invariants of the time-reversal operator:
  Theory and experiment}, J. Acoust. Soc. Am. 120 (2006), pp. 875--883.

\bibitem[11]{borcea}
L. Borcea, G. Papanicolaou, and C. Tsogka, {\itshape Adaptative interferometric
  imaging in clutter and optimal illumination}, Inverse Problems 22 (2006), pp.
  1405--1436.

\bibitem[12]{mordant}
N. Mordant, C. Prada, and M. Fink, {\itshape Highly resolved detection and
  selective focusing in a waveguide using the d.o.r.t method}, J. Acoust. Soc.
  Am. 105 (1999), pp. 2634--2642.

\bibitem[13]{lingevitch}
J.F. Lingevitch, H.C. Song, and W.A. Kuperman, {\itshape Time reversed
  reverberation focusing in a waveguide}, J. Acoust. Soc. Am. 111 (2002), pp.
  2609--2614.

\bibitem[14]{Carin04}
L. Carin, H. Liu, T. Yoder, L. Couchman, B. Houston, and J. Bucaro, {\itshape
  Wideband time-reversal imaging of an elastic target in an acoustic
  waveguide}, J. Acoust. Soc. Am. 115 (2004), pp. 259--268.

\bibitem[15]{kerbrat}
E. Kerbrat, C. Prada, D. Cassereau, and M. Fink, {\itshape Imaging the presence
  of grain noise using the decomposition of the time reversal operator}, J.
  Acoust. Soc. Am. 113 (2003), pp. 1230--1240.

\bibitem[16]{gaumond}
C. Gaumond, D. Fromm, J. Lingevitch, R. Menis, G. Edelmann, D. Calvo, and E.
  Kim, {\itshape Demonstration at sea of the
  decomposition-of-the-time-reversal-operator technique}, J. Acoust. Soc. Am.
  119 (2006), pp. 976--990.

\bibitem[17]{prada3}
C. Prada, J. de Rosny, D. Clorennec, J.G. Minonzio, A. Aubry, M. Fink, L.
  Berniere, P. Billand, S. Hibral, and T. Folegot, {\itshape Experimental
  detection and focusing in shallow water by decomposition of the time reversal
  operator}, J. Acoust. Soc. Am. 122 (2007), pp. 761--768.

\bibitem[18]{Saillard99}
H. Tortel, G. Micolau, and M. Saillard, {\itshape Decomposition of the time
  reversal operator for electromagnetic scattering}, J. Electromagn. Waves
  Appl. 13 (1999), pp. 687--719.

\bibitem[19]{Micolau03}
G. Micolau, M. Saillard, and P. Borderies, {\itshape DORT method as applied to
  ultrawideband signals for detection of buried objects}, IEEE Trans. Geosci.
  Remote Sens. 41 (2003), pp. 1813--1820.

\bibitem[20]{iakoleva2}
E. Iakovleva, S. Gdoura, D. Lesselier, and G. Perrusson, {\itshape Multi-static
  response matrix of a 3-{D} inclusion in a half space and {MUSIC} imaging},
  IEEE Trans. Antennas Propagat. 55 (2007), pp. 2598--2609.

\bibitem[21]{iakoleva}
E. Iakoleva and D. Lesselier, {\itshape Multistatic response matrix of
  spherical scatterers and the back-propagation of singular fields}, IEEE
  Trans. Antennas Propagat. 56 (2008), pp. 825--833.

\bibitem[22]{badereau}
D. {de Badereau}, H. Roussel, and W. Tabbara, {\itshape Radar remote sensing of
  forest at low frequencies: a two dimensional full wave approach}, J.
  Electromagn. Waves Applic. 17 (2003), pp. 921--949.

\bibitem[23]{nguyen}
H. Nguyen, H. Roussel, and W. Tabbara, {\itshape A coherent model of forest
  scattering and {SAR} imaging in the {VHF} and {UHF} band}, IEEE Trans.
  Geosci. Remote Sens. 44 (2006), pp. 838--848.

\bibitem[24]{tabbara}
Y. Ziad\'{e}, H. Roussel, M. Lesturgie, and W. Tabbara, {\itshape A coherent
  model of forest propagation - {A}pplication to detection and localisation of
  targets using the {DORT} method}, IEEE Trans. Antennas Propagat. 56 (2008),
  pp. 1048--1057.

\bibitem[25]{brody}
T. Brody, J. Flores, J. Franch, P. Mello, A. Pandey, and S. Song, {\itshape
  Random-matrix Physics: spectrum and strength fluctuations}, Rev. Mod. Phys.
  53 (1981), pp. 385--479.

\bibitem[26]{ellegard}
C. Ellegaard, T. Guhr, K. Lindemann, J. Nygard, and M. Oxborrow, {\itshape
  Symmetry breaking and spectral statistics of acoustic resonances in quartz
  blocks}, Phys. Rev. Lett. 77 (2003), pp. 4918--4921.

\bibitem[27]{cun}
Y.L. Cun, I. Kanter, and S.A. Solla, {\itshape Eigenvalues of covariance
  matrices: Application to neural-network learning}, Phys. Rev. Lett. 66
  (1991), pp. 2396--2399.

\bibitem[28]{laloux}
L. Laloux, P. Cizeau, J-P. Bouchaud, and M. Potters, {\itshape Noise Dressing
  of Financial Correlation Matrices}, Phys. Rev. Lett. 83 (1999), pp.
  1467--1470.

\bibitem[29]{sengupta}
A.M. Sengupta and P.P. Mitra, {\itshape Distribution of singular values for
  some random matrices}, Phys. Rev. E 60 (1999), pp. 3389--3392.

\bibitem[30]{johnstone}
I. Johnstone, {\itshape On the distribution of the largest eigenvalue in
  principal components analysis}, Ann. Statist. 29 (2001), pp. 295--327.

\bibitem[31]{karoui}
N. El-Karoui, {\itshape Spectrum estimation for large dimensional covariance
  matrices using random matrix theory}, Ann. Statist. 36 (2008), pp.
  2757--2790.

\bibitem[32]{aubry09}
A. Aubry and A. Derode, {\itshape Random matrix theory applied to acoustic
  backscattering and imaging in complex media}, Phys. Rev. Lett. 102 (2009), p.
  084301.

\bibitem[33]{marcenko}
V. Mar\u{c}enko and L. Pastur, {\itshape Distributions of eigenvalues for some
  sets of random matrices}, Math. USSR-Sbornik 1 (1967), pp. 457--483.

\bibitem[34]{derode4}
A. Derode, V. Mamou, and A. Tourin, {\itshape Influence of correlations between
  scatterers on the attenuation of the coherent wave in a random medium}, Phys.
  Rev. E 74 (2006), p. 036606.

\bibitem[35]{Ishimaru1}
Y. Kuga and A. Ishimaru, {\itshape Retroreflectance from a dense distribution
  of spherical particles}, J. Opt. Soc. Am. A 1 (1984), p. 831.

\bibitem[36]{akkermans}
E. Akkermans, P-E. Wolf, and R. Maynard, {\itshape Coherent backscattering of
  light by disordered media: Analysis of the peak line shape}, Phys. Rev. Lett.
  56 (1986), pp. 1471--1474.

\bibitem[37]{wolf}
P-E. Wolf and G. Maret, {\itshape Weak localization and coherent backscattering
  of photons in disordered Media}, Phys. Rev. Lett. 55 (1985), pp. 2696--2699.

\bibitem[38]{tourin}
A. Tourin, A. Derode, P. Roux, B.A. {van Tiggelen}, and M. Fink, {\itshape
  Time-dependent backscattering of acoustic waves}, Phys. Rev. Lett. 79 (1997),
  pp. 3637--3639.

\bibitem[39]{tourin2}
A. Tourin, A. Derode, A. Peyre, and M. Fink, {\itshape Transport parameters for
  an ultrasonic pulsed wave propagating in a multiple scattering medium}, J.
  Acoust. Soc. Am. 108 (2000), pp. 503--512.

\bibitem[40]{aubry}
A. Aubry and A. Derode, {\itshape Ultrasonic imaging of highly scattering media
  from local measurements of the diffusion constant: separation of coherent and
  incoherent intensities.}, Phys. Rev. E 75 (2007), p. 026602.

\bibitem[41]{ishimaru}
A. Ishimaru {\itshape Wave Propagation and Scattering in Random Media},
  Academic Press, New York, 1978.

\bibitem[42]{mamou}
V. Mamou, {\itshape Caract\'erisation ultrasonore d'\'echantillons
  h\'et\'erog\`enes multiplement diffuseurs}, Universit\'e Paris 7 - Denis
  Diderot, 2005, http://tel.archives-ouvertes.fr.

\bibitem[43]{goodman}
J.W. Goodman {\itshape Statistical Optics},    Wiley \& Sons, New York, 1985.

\bibitem[44]{derode3}
A. Derode and M. Fink, {\itshape Partial coherence of transient ultrasonic
  fields in anisotropic random media: Application to coherent echo detection},
  J. Acoust. Soc. Am. 101 (1997), pp. 690--704.

\bibitem[45]{wiersma}
D.S. Wiersma, M.P. {van Albada}, B.A. {van Tiggelen}, and A. Lagendijk,
  {\itshape Experimental evidence for recurrent multiple scattering events of
  light in disordered media}, Phys. Rev. Lett. 74 (1995), pp. 4193--4196.

\bibitem[46]{Nieuwenhuizen}
T.M. Nieuwenhuizen and M.C.W. {van Rossum}, {\itshape Intensity distributions
  of waves transmitted through a multiple scattering medium}, Phys. Rev. Lett.
  74 (1995), pp. 2674--2677.

\bibitem[47]{hu}
H. Hu, A. Strybulevych, J.H. Page, S.E. Skipetrov, and B.A. {van Tiggelen},
  {\itshape Localization of ultrasound in a three-dimensional elastic network},
  Nature Phys. 4 (2008), pp. 945--948.

\bibitem[48]{haney}
M. Haney and R. Snieder, {\itshape Breakdown of wave diffusion in 2D due to
  Loops}, Phys. Rev. Lett. 91 (2003), p. 093902.

\bibitem[49]{bryc}
W. Bryc, A. Dembo, and T. Jiang, {\itshape Spectral measure of large random
  Hankel, Markov and Toeplitz matrices}, Ann. Probab. 34 (2006), pp. 1--38.

\bibitem[50]{chambers}
D. Chambers and A. Gautesen, {\itshape Time reversal for a single spherical
  scatterer}, J. Acoust. Soc. Am. 109 (2001), pp. 2616--2624.

\bibitem[51]{minonzio2}
J-G. Minonzio, C. Prada, D. Chambers, and M. Fink, {\itshape Characterization
  of subwavelength elastic cylinders with the decomposition of the
  time-reversal operator}, J. Acoust. Soc. Am. 117 (2005), pp. 789--798.

\bibitem[52]{zhao}
H. Zhao, {\itshape Analysis of the response matrix for an extended target},
  SIAM J. Appl. Math. 64 (2004), pp. 725--745.

\bibitem[53]{aubry06}
A. Aubry, J. {de Rosny}, J-G. Minonzio, C. Prada, and M. Fink, {\itshape
  Gaussian beams and Legendre polynomials as invariants of the time reversal
  operator for a large rigid cylinder}, J. Acoust. Soc. Am. 120 (2006), pp.
  2746--2754.

\bibitem[54]{robert}
J.L. Robert and M. Fink, {\itshape The prolate spheroidal wave functions as
  invariants of the time reversal operator for an extended scatterer in the
  Fraunhofer approximation}, J. Acoust. Soc. Am. 125 (2009), pp. 218--226.

\bibitem[55]{mehta}
M. Mehta {\itshape Random Matrices},    Academic Press, Boston, MA, 1991.

\bibitem[56]{pastur}
L. Pastur, {\itshape On the universality of the level spacing distribution for
  some ensembles of random matrices}, Lett. Math. Phys. 25 (1992), pp.
  259--265.

\bibitem[57]{tracy2}
C.A. Tracy and H. Widom, {\itshape Level-spacing distributions and the {A}iry
  kernel}, Commun. Math. Phys. 159 (1994), pp. 151--174.

\bibitem[58]{tracy}
---{}---{}---, {\itshape On orthogonal and symplectic matrix ensembles},
  Commun. Math. Phys. 177 (1996), pp. 727--754.

\bibitem[59]{johansson}
K. Johansson, {\itshape Shape fluctuations and random matrices}, Commun. Math.
  Phys. 209 (2000), pp. 437--476.

\bibitem[60]{karoui2}
N. El-Karoui, {\itshape Recent results about the largest eigenvalue of random
  covariance matrices and statistical application}, Acta Phys. Polon. B 36
  (2005), pp. 2681--2697.

\bibitem[61]{meckes}
M. Meckes, {\itshape On the spectral norm of a random Toeplitz matrix},
  Electron. Commun. Probab. 12 (2007), pp. 315--325.

\bibitem[62]{adamczak}
R. Adamczak, {\itshape A few remarks on the operator norm of random {T}oeplitz
  matrices}, J. Theor. Probab  (2009).

\bibitem[63]{chatterjee}
S. Chatterjee, {\itshape Fluctuations of eigenvalues and second-order
  {P}oincar\'{e} inequalities}, Probab. Theory Related Fields 143 (2009), pp.
  1--40.

\bibitem[64]{aubry09_2}
A. Aubry and A. Derode, {\itshape Detection and imaging in a random medium: a
  matrix method to overcome multiple scattering and aberration}, J. Appl. Phys. 106 (2009), p. 044903.

\end{thebibliography}

\appendix

\section{\label{app:temp_window}}

The time-window length $\Delta t$ is chosen so that signals associated with the same scattering path within the medium are in the same time window. For the sake of simplicity, here we only deal with single scattering paths but we would obtain the same results if multiple-scattering paths were considered. 

To calculate $\Delta t$, we have to find the two single-scattering paths, associated to the same scattering event, for which the difference of travel length is the highest. To that aim, the directivity of transducers has to be taken into account. The major part of the energy is transmitted (received) towards (from) scatterers located into a cone whose aperture is $2 \theta_{max}$. When the response $k_{ij}$ is considered, only the scatterers contained inside the volume common to the directivity cones of elements $i$ and $j$ have to be considered. 

At each depth $Z_s$, we have to optimize the lateral position $X_s$ of the scatterer which results in the largest difference of travel length between two single-scattering paths. These paths are associated with two source/receiver couples $(i,j)$ and $(l,m)$. Thus, we have to optimize simultaneously the position $X_s$, and the couples $(i,j)$ and $(l,m)$. This issue can be simplified because we have to deal with single-scattering paths. Because of the equivalence of the forward and return waves, source and receiver of each couples are in fact identical. So, we have to find the elements $i$ and $l$ which are respectively the furthest and the nearest elements from the scatterer whose coordinates are $(X_s,Z_s)$, $X_s$ remaining unknown. The result of this optimization differs according to the depth $Z_s$ of the scatterer :
\begin{figure}[htbp] 
\begin{center}
\includegraphics{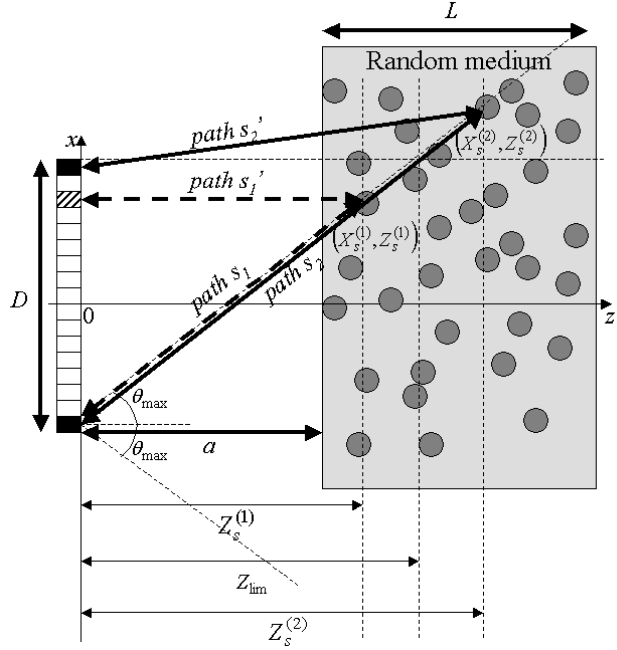}
\caption{\label{fig:fig9}Choice of the appropriate time-window length $\Delta t$. The single scattered paths $s$ and $s'$ are associated with the reflection on the same scatterer. Two situations can occur. The case $z=Z_s^{(1)}$, when $Z_s^{(1)}<Z_{lim}$, corresponds to the dashed arrows. The two paths $s_1$ and $s'_1$ have been chosen so that the difference of travel length is the largest. In the case $z=Z_s^{(2)}$(with $Z_s^{(2)}>Z_{lim}$ ), continuous arrows represent the two paths $s_2$ and $s'_2$ for which the difference of travel length is the largest.}
\end{center}
\end{figure}
\begin{itemize}
\item $Z_s<Z_{lim}=\frac{D}{2} \cos \theta_{max}$: the two paths $s_1$ and $s'_1$ corresponding to the largest difference of travel length are shown with dashed arrows in Fig.\ref{fig:fig9}. $s_1$ is the longest path and contributes to the signal $k_{11}$. It is linked with a scatterer which is located on the top generatrix of the directivity cone of the first transducer. $s'_1$ is the shortest path. It is associated with the $p^{th}$ element of the array and thus contributes to the signal $k_{pp}$. The transducer $p$ is at the same transverse position $X_s$ as the scatterer. So, for $Z_s<Z_{lim}$, $\Delta t$ is then given by
\begin{equation}
\label{eqn:app_1}
\Delta t =\frac{2Z_s}{c}\left [ \frac{1}{\cos \theta_{max}} - 1 \right ]
\end{equation}
\item $Z_s>Z_{lim}$: the two paths $s_2$ and $s'_2$ are depicted with continuous arrows in Fig.\ref{fig:fig9}. As previously, $s_2$ is the longest path and contributes to the signal $k_{11}$. It is still linked with a scatterer which is located on the top generatrix of the directivity cone of the first transducer. $s'_2$ is the shortest path. It is associated with the $N^{th}$ element of the array and thus contributes to the signal $k_{NN}$. For $Z_s>Z_{lim}$, $\Delta t$ is hence given by
\begin{equation}
\label{eqn:app_2}
\Delta t =\frac{2}{c}\left [ \frac{Z_s}{\cos \theta_{max}} - \sqrt{\left  (Z_s\tan \theta_{max} + D \right)^2+Z_s^2 } \right ]
\end{equation}
where $D$ is the array aperture. When $Z_s \rightarrow \infty $, Eq.\ref{eqn:app_2} becomes
\begin{equation}
\label{eqn:app_3}
\lim_{Z_s \rightarrow +\infty} \Delta t = \frac{2D \sin \theta_{max}}{c}
\end{equation}
\end{itemize}
In practice, we have to assess $\Delta t$ by considering the maximum depth $Z_{max}=a+L$. For the experiment described in Sec.\ref{sec:ms}, $Z_{max}=65$ mm. Considering an aperture angle $\theta_{max}=27.5 \deg$, we obtain a value $\Delta t \simeq 10 \mu s $. For the experiment described in Sec.\ref{sec:ss}, $Z_{max}=150$ mm. Thus the value $\Delta t$ should be fixed to $14 \mu s$. Nevertheless, given the finite width of the agar gel sample (8 cm), there is no need to use such long temporal windows and a time-window length $\Delta t = 10 \mu s$ is also considered.

\section{\label{app:var_double}}
As seen in Sec.\ref{subsec:identical}, the coherent backscattering effect arises in the diagonal of matrix $\tilde{\mathbf{K}}$: the variance of coefficients $\tilde{k}_{ii}$ is twice that of off-diagonal elements. We want to estimate the influence of the coherent backscattering effect on the singular values distribution. To that aim, we will investigate its influence on the statistical properties of the autocorrelation matrix.

Let us first consider a random matrix $\mathbf{R}$ of dimension $N \times N$. We assume that the coefficients of the matrix $\mathbf{R}$ are complex gaussian random variables \textit{i.i.d}, with a zero-mean and a variance of $1/N$. The theoretical singular values distribution of a random matrix $\mathbf{R}$ (the quarter circle law) is deduced directly from the eigenvalues distribution of the autocorrelation matrix $\mathbf{A}=\mathbf{RR}^{\dag}$ (the so-called Mar\u{c}enko-Pastur law \cite{tulino}), since singular values of $\mathbf{R}$ correspond to the square root of eigenvalues of $\mathbf{A}=\mathbf{RR}^{\dag}$. We now focus on the statistical properties of $\mathbf{A}$.  
The entries $a_{lm}$ of matrix $\mathbf{A}$ are given by
\begin{equation}
\label{eqn:alm}
a_{lm}=\sum_{p=1}^Nr_{lp}r_{mp}^*
\end{equation}
Let us calculate the mean and the variance of coefficients $a_{lm}$. $\left < a_{lm} \right >$ is given by
\begin{eqnarray}
\left < a_{lm} \right >& =& \sum_{p=1}^N \left <r_{lp}r_{mp}^*\right >\nonumber \\
\left < a_{lm} \right >& =&  \left \{ 
\begin{array}{cl}
0 &  \mbox{if }l\neq m \\
\sum_{p=1}^N \left < \left | r_{lp} \right |^2 \right > = 1 & \mbox{otherwise} 
\end{array}
\right. \nonumber
\end{eqnarray}
The mean of coefficients $a_{lm}$ is not nil only for diagonal elements
\begin{equation}
\left < a_{lm} \right > =  \delta_{lm}
\label{eqn:mean_ar}
\end{equation}
$\left < \left | a_{lm} \right |^2 \right >$ can be developed as
\begin{equation*}
\left < \left | a_{lm} \right |^2 \right >  = \sum_{p=1}^N \sum_{q=1}^N \left < r_{lp}  r_{mp}^*  r_{lq}^*  r_{mq} \right >
\end{equation*}
Using the moment theorem, $\left < \left | a_{lm} \right |^2 \right >$ becomes
\begin{equation}
\label{eqn:alm_s}
\left < \left | a_{lm} \right |^2 \right >  = \underbrace{\sum_{p=1}^N \sum_{q=1}^N \left < r_{lp}  r_{mp}^* \right> \left <  r_{lq}^*  r_{mq} \right >}_{\left | \left< a_{lm} \right > \right |^2} + \underbrace {\sum_{p=1}^N \sum_{q=1}^N \left < r_{lp}  r_{lq}^* \right> \left <  r_{mp}^*  r_{mq} \right >}_{\mbox{var} \left [a_{lm} \right] }
\end{equation}
To calculate $\mbox{var} \left [a_{lm} \right] $, we use the fact that
\begin{equation*}
\left < r_{lp}  r_{lq}^* \right > =\frac{1}{N} \delta_{pq}
\end{equation*}
We obtain 
\begin{equation}
\mbox{var} \left [a_{lm} \right] = \frac{1}{N^2}\sum_{p=1}^N \sum_{q=1}^N \delta_{pq} =\frac{1}{N} 
\label{eqn:var_ar}
\end{equation}
Finally, the off-diagonal entries of $\mathbf{A}$ are complex random variables with zero mean and variance $1/N$. The diagonal elements of $\mathbf{A}$ are also complex random variables but with a mean equal to 1 and a variance of $1/N$.

Now that we have calculated the mean and variance of the autocorrelation coefficients $a_{lm}$ built from a ``classical'' random matrix $\mathbf{R}$, we focus on the effect of coherent backscattering. To that aim, we consider the matrix $\mathbf{\tilde{K}}$ which is obtained experimentally in the multiple scattering regime. This matrix is built from the array response matrix $\mathbf{K}$:
\begin{equation}
\tilde{k}_{ij}=\frac{k_{ij}}{\sqrt{\frac{1}{N}\sum_{p=1}^N\sum_{q=1}^N \left | k_{pq} \right |^2 }}
\label{eqn:tilde_K}
\end{equation}
Eq.\ref{eqn:tilde_K} is another expression for the renormalization presented in Eq.\ref{eqn:eq2}. By construction, we have
\begin{equation}
\frac{1}{N}\sum_{i=1}^N\sum_{j=1}^N \left | \tilde{k}_{ij} \right |^2=1
\label{eqn:constraint}
\end{equation}
Note that the random matrix $\mathbf{R}$ verifies the same property. But, as pointed out in Sec.\ref{subsec:identical}, the variance of elements of $\mathbf{\tilde{K}}$ is not constant. In particular, the matrix $\mathbf{\tilde{K}}$ exhibits a doubled variance along its diagonal because of the coherent backscattering phenomenon. To estimate its effect on the singular value spectrum, we model $\mathbf{\tilde{K}}$ as a matrix whose coefficients $\tilde{k}_{ij}$ are complex gaussian random variables independently distributed with mean zero and a variance $\sigma_{ij}^2$ defined as
\begin{equation}
\sigma_{ij}^2= \left \{
\begin{array}{rl}
\sigma^2 & \mbox{if }i\neq j \\
2\sigma^2 & \mbox{if }i=j 
\end{array}
\label{eqn:sigma_ij}
\right.
\end{equation}
Eq.\ref{eqn:constraint} leads to
\begin{eqnarray*}
\frac{1}{N^2}\sum_{p=1}^N\sum_{q=1}^N \sigma_{pq}^2=\frac{1}{N} \\
\frac{1}{N^2}\sum_{p=1}^N \left [ (N-1)\sigma^2+ 2 \sigma^2\right]=\frac{1}{N}\\
\frac{N(N+1)}{N^2} \sigma^2=\frac{1}{N} \mbox{,}
\end{eqnarray*}
which yields
\begin{equation}
\sigma^2=\frac{1}{N+1}
\label{eqn:sigma}
\end{equation}
$\sigma_{ij}^2$ is finally given by
\begin{equation}
\sigma_{ij}^2= \frac{1+\delta_{ij}}{N+1}
\end{equation}
The variance of elements $\tilde{k}_{ij}$ is equal to $\frac{1}{N+1}$ for off-diagonal elements ($i \neq j$) and twice for diagonal elements ($i=j$).

Let us focus now on the autocorrelation matrix $\mathbf{B}=\mathbf{\tilde{K}}\mathbf{\tilde{K}}^{\dag}$. Its coefficients are defined as
\begin{equation}
b_{lm}=\sum_{p=1}^N\tilde{k}_{lp}\tilde{k}_{mp}^*
\end{equation}
The mean of coefficients $b_{lm}$ is given by
\begin{eqnarray*}
\left < b_{lm} \right >& =& \sum_{p=1}^N \left <\tilde{k}_{lp}\tilde{k}_{mp}^*\right >\\
\left < b_{lm} \right >& =& 
\left \{ 
\begin{array}{l}
0  \mbox{, if }l\neq m \\
\sum_{p=1}^N \left < \left | \tilde{k}_{lp} \right |^2 \right > = \sigma_{ll}^2 + \sum_{p \neq l}^N \sigma_{lp}^2 = \frac{2}{N+1}+\frac{N-1}{N+1} =1  \mbox{, otherwise}
\end{array}
\right.
\end{eqnarray*}
We obtain the same mean for $b_{lm}$ as for $a_{lm}$ (Eq.\ref{eqn:mean_ar})
\begin{equation}
\left < b_{lm} \right >= \delta_{lm}
\label{eqn:mean_b}
\end{equation}
As to the variance of coefficient $b_{lm}$, it can be expressed as
\begin{eqnarray}
\mbox{var} \left [b_{lm} \right]& = &\sum_{p=1}^N \sum_{q=1}^N \underbrace {\left < \tilde{k}_{lp}  \tilde{k}_{lq}^* \right> }_{\sigma_{lq}^2\delta_{pq}}\underbrace{\left <  \tilde{k}_{mp}^*  \tilde{k}_{mq} \right >}_{\sigma_{mq}^2\delta_{pq}} \nonumber\\
\mbox{var} \left [b_{lm} \right]& = & \sum_{q=1}^N \sigma_{lq}^2 \sigma_{mq}^2 \nonumber\\
\mbox{var} \left [b_{lm} \right]& = & \frac{1}{(N+1)^2}\sum_{q=1}^N (1+\delta_{lq})(1+\delta_{mq}) \nonumber \\
\mbox{var} \left [b_{lm} \right]& = & \frac{1}{(N+1)^2}(N+2+\delta_{lm}) \nonumber \\
\mbox{var} \left [b_{lm} \right]& = & \frac{N+2+\delta_{lm}}{(N+1)^2}
\label{eqn:var_blm}
\end{eqnarray}
The statistical properties of the autocorrelation matrices $\mathbf{A}$ and $\mathbf{B}$ are summed up in Tab.\ref{tab:tab_recap}
\begin{table}[htbp]
\begin{center}
{\begin{tabular}{|c|c|}
\hline
``Classical'' random matrix $\mathbf{R}$ & Experimental matrix $\mathbf{\tilde{K}}$ \\
 $\mathbf{A}=\mathbf{R}\mathbf{R}^{\dag}$ & $\mathbf{B}=\mathbf{\tilde{K}}\mathbf{\tilde{K}}^{\dag}$\\
\hline
$\left < a_{lm} \right >= \delta_{lm}$ & $\left < b_{lm} \right >= \delta_{lm}$ \\
\hline
$\mbox{var} \left [a_{lm} \right] = \frac{1}{N}$ & $\mbox{var} \left [b_{lm} \right] =  \frac{N+2+\delta_{lm}}{(N+1)^2}$ \\
\hline
\end{tabular}}
\caption{\label{tab:tab_recap}Statistical properties of autocorrelation matrices $\mathbf{A}$ and $\mathbf{B}$, depending on whether the initial matrix exhibits the coherent backscattering effect or not.}
\end{center}
\end{table}
The coherent backscattering effect increases the variance of the autocorrelation coefficients, compared to the \textit{classical case}. Thus, the eigenvalues spectrum of the autocorrelation matrix and also the singular values distribution of $\mathbf{\tilde{K}}$ should be modified by the coherent backscattering effect. But, for $N>>1$, the difference between $\frac{1}{N}$, $\frac{N+2}{(N+1)^2}$ and $\frac{N+3}{(N+1)^2}$ becomes negligible. For instance, the relative error is about 3\% for $N=32$. So, the doubled variance of diagonal elements of $\mathbf{\tilde{K}}$ does not perturb significantly the singular values distribution. 

\section{\label{app:sym}}

We investigate the influence of reciprocity on the statistical properties of the autocorrelation matrix. We will consider first the case of a ``classical'' random matrix and study the statistical properties of its autocorrelation matrix. The case of a symetric random matrix will be studied afterwards. The comparison of the results obtained in both cases will allow us to explain and quantify the influence of symmetry on the singular values distribution. As we will see, the symmetry implies correlations between the diagonal elements of the autocorrelation matrix.

Let us first consider a random matrix $\mathbf{R}$ of dimension $N \times N$. We assume that the coefficients of the matrix $\mathbf{R}$ are complex gaussian random variables \textit{i.i.d}, zero-mean and with variance $1/N$. The entries $a_{lm}$ of the autocorrelation matrix $\mathbf{A}=\mathbf{R}\mathbf{R}^{\dag}$ are given by
\begin{equation*}
a_{lm}=\sum_{p=1}^Nr_{lp}r_{mp}^*
\end{equation*}
Let us calculate the correlation coefficient $\Theta^{A}_{lm}$ between diagonal elements $a_{ll}$ and $a_{mm}$, which is defined as
\begin{equation}
\label{eqn:corr_all}
\Theta^{A}_{lm}=\frac{\left < a_{ll} a_{mm}^* \right > -\left | \left < a_{ll}  \right> \right |^2  }{\mbox{var} \left [a_{ll} \right ]}
\end{equation}
The correlation term $\left < a_{ll} a_{mm}^* \right >$ can be developed as
\begin{equation*}
\left < a_{ll} a_{mm}^* \right >  = \sum_{p=1}^N \sum_{q=1}^M \left < r_{lp} r_{lp}^* r_{mq} r_{mq}^* \right >
\end{equation*}
Using the moment theorem, the last equation becomes
\begin{equation}
\label{eqn:w1}
\left < a_{ll} a_{mm}^* \right >  = \underbrace {\sum_{p=1}^N \sum_{q=1}^M \left < r_{lp} r_{lp}^* \right> \left< r_{mq} r_{mq}^* \right >}_{\left | \left < a_{ll}  \right> \right |^2 } + \sum_{p=1}^N \sum_{q=1}^M \left < r_{lp} r_{mq}^* \right> \left< r_{mq} r_{lp}^* \right > 
\end{equation}
The second sum can be calculated using the fact that
\begin{equation}
\label{eqn:w2}
\left < r_{lp} r_{mq}^* \right> = \frac{\delta_{lm}\delta_{pq}}{N}
\end{equation}
which yields
\begin{eqnarray*}
\left < a_{ll} a_{mm}^* \right > - \left | \left < a_{ll}  \right> \right |^2  &= &\sum_{p=1}^N \sum_{q=1}^M \frac{\delta_{lm}\delta_{pq}}{N^2} \\
\left < a_{ll} a_{mm}^* \right > - \left | \left < a_{ll}  \right> \right |^2  &= & \frac{\delta_{lm}}{N}
\end{eqnarray*}
Using the fact that $\mbox{var} \left [a_{ll} \right]=\frac{1}{N}$ (Eq.\ref{eqn:var_ar}), the correlation coefficient $\Theta^{A}_{lm}$(Eq.\ref{eqn:corr_all}) is given by
\begin{equation}
\label{eqn:corr_all_result}
\Theta^{A}_{lm}= \delta_{lm}
\end{equation}
This last equation means that the diagonal elements $a_{ll}$ of $\mathbf{A}$ are totally decorrelated from each other.

We now consider the case of a random but symmetric matrix like $\mathbf{\tilde{K}}$. Its coefficients are complex gaussian random variables identically distributed, with mean zero and variance $1/N$. The only difference with a ``classical'' random matrix $\mathbf{R}$ is the fact that $\tilde{k}_{ij}=\tilde{k}_{ji}$. We neglect here the coherent backscattering effect(see Appendix \ref{app:var_double}) in order to focus on the influence of reciprocity. 

Let us study the statistical properties of the autocorrelation matrix $\mathbf{B}=\mathbf{\tilde{K}}\mathbf{\tilde{K}}^{\dag}$. One can show that symmetry has no effect on the mean and the variance of $b_{lm}$
\begin{eqnarray*}
\left < b_{lm}\right > \equiv \left < a_{lm}\right > & = & \frac{\delta_{lm}}{N} \\
\mbox{var} \left [ b_{lm} \right] \equiv \mbox{var} \left [ a_{lm} \right] &=& \frac{1}{N}
\end{eqnarray*}
As done previously for the matrix $\mathbf{A}$, we define the correlation coefficient $\Theta^{B}_{lm}$ between diagonal elements $b_{ll}$ and $b_{mm}$
\begin{equation}
\label{eqn:corr_all_B}
\Theta^B_{lm}=\frac{\left < b_{ll} b_{mm}^* \right > -\left | \left < b_{ll}  \right> \right |^2  }{\mbox{var} \left [b_{ll} \right ]}
\end{equation}
The correlation term $\left < b_{ll} b_{mm}^* \right >$ can be developed as
\begin{equation*}
\left < b_{ll} b_{mm}^* \right >  = \sum_{p=1}^N \sum_{q=1}^M \left < \tilde{k}_{lp} \tilde{k}_{lp}^* \tilde{k}_{mq} \tilde{k}_{mq}^* \right >
\end{equation*}
Using the moment theorem, the last equation becomes
\begin{equation*}
\left < b_{ll} b_{mm}^* \right >  = \underbrace {\sum_{p=1}^N \sum_{q=1}^M \left < \tilde{k}_{lp} \tilde{k}_{lp}^* \right> \left< \tilde{k}_{mq} \tilde{k}_{mq}^* \right >}_{\left | \left < b_{ll}  \right> \right |^2 } + \sum_{p=1}^N \sum_{q=1}^M \left < \tilde{k}_{lp} \tilde{k}_{mq}^* \right> \left< \tilde{k}_{mq} \tilde{k}_{lp}^* \right > 
\end{equation*}
But, because $\tilde{k}_{lp}=\tilde{k}_{pl}$, we have
 \begin{equation}
\label{eqn:w3}
\left < \tilde{k}_{lp} \tilde{k}_{mq}^* \right> = \frac{\delta_{lm}\delta_{pq}}{N} +\frac{\delta_{lq}\delta_{mp}}{N}
\end{equation}
We obtain:
\begin{equation*}
\left < b_{ll} b_{mm}^* \right > - \left | \left < b_{ll}  \right> \right |^2  =  \frac{\delta_{lm}}{N} + \frac{1}{N^2}
\end{equation*}
Using the fact that $\mbox{var} \left [b_{ll} \right]=\frac{1}{N}$, the correlation coefficient $\Theta^{B}_{lm}$(Eq.\ref{eqn:corr_all_B}) is finally given by
\begin{equation}
\label{eqn:corr_all_result_2}
\Theta^{B}_{lm}= \delta_{lm} + \frac{1}{N}
\end{equation}
This last equation means that a correlation exists between the diagonal elements $b_{ll}$ of $\mathbf{B}$, due to reciprocity. 
The correlations between diagonal elements of matrices $\mathbf{A}$ and $\mathbf{B}$ are compared in Tab.\ref{tab:tab_recap2}
\begin{table}[htbp]
\begin{center}
{\begin{tabular}{|c|c|}
\hline
``Classical'' random matrix $\mathbf{R}$ & Symmetric random matrix $\mathbf{\tilde{K}}$ \\
 $\mathbf{A}=\mathbf{R}\mathbf{R}^{\dag}$ & $\mathbf{B}=\mathbf{\tilde{K}}\mathbf{\tilde{K}}^{\dag}$\\
\hline
$\Theta^{A}_{lm}= \delta_{lm}$ & $\Theta^{B}_{lm}= \delta_{lm} + \frac{1}{N}$ \\
\hline
\end{tabular}}
\caption{\label{tab:tab_recap2}Correlation between diagonal elements of autocorrelation matrices $\mathbf{A}$ and $\mathbf{B}$, depending on whether the initial matrix is symmetric or not.}
\end{center}
\end{table}
Even if this correlation is not too large ($1/N \simeq 3$\% with $N=32$), it has an influence on the eigenvalue spectrum of the autocorrelation matrix, hence on the singular values distribution of the matrix $\mathbf{\tilde{K}}$. This explains partly the deviation from the quarter circle law, which we pointed out numerically (see Fig.\ref{fig:fig5}(b)). Note that for $N>>1$, the influence of symmetry should vanish.

\section{\label{app:sing_value}}
We assume that the matrix $\mathbf{K^T}$ associated to the target is of rank 1. $\lambda_1^T$ is the only non-zero singular value of $\mathbf{K^T}$. The trace of the autocorrelation matrix $\mathbf{K^T}\mathbf{K^T}^{\dag}$ can then be expressed as
\begin{equation}
\label{eqn:trace}
\mbox{Trace}\left [ \mathbf{K^T}\mathbf{K^T}^{\dag}\right ]= \sum_{p=1}^N \left (\lambda_p^T\right)^{2} = \left (\lambda_1^T\right)^{2}
\end{equation}
$\sigma_T^2$ is the power of signals associated with the target : $$\sigma_T^2= \frac{1}{N^2}\sum_{i=1}^N\sum_{j=1}^N  \left < \left |k^T_{ij} \right |^2 \right > \mbox{.}$$
The trace of $\mathbf{K^T}\mathbf{K^T}^{\dag}$ can also be expressed as
\begin{eqnarray}
\mbox{Trace}\left [ \mathbf{K^T}\mathbf{K^T}^{\dag}\right ] &=& \sum_{i=1}^N\sum_{j=1}^N \left |k^T_{ij} \right |^2  \nonumber \\
\mbox{Trace}\left [ \mathbf{K^T}\mathbf{K^T}^{\dag}\right ] &=& N^2 \sigma_T^2
\label{eqn:trace2}
\end{eqnarray}
From Eq.\ref{eqn:trace} and Eq.\ref{eqn:trace2}, we deduce
\begin{equation}
\label{eqn:trace3}
\lambda_1^T = N\sigma_T
\end{equation}

Now, we consider the case of the matrix $\mathbf{K}$. This matrix is the sum of two contributions (Eq.\ref{eqn:decompos_K}) :
\begin{itemize}
\item $\mathbf{K^T}$ which corresponds to the direct echo reflected by the target.
\item $\mathbf{K^R}$ which corresponds to the response of the random medium (or to additive noise).
\end{itemize}
The trace of the autocorrelation matrix $\mathbf{K}\mathbf{K^{\dag}}$ is given by
\begin{equation}
\label{eqn:trace4}
\mbox{Trace}\left [ \mathbf{K}\mathbf{K}^{\dag}\right ]= \sum_{p=1}^N \lambda_p^2
\end{equation}
We note $\sigma_R^2$ the mean power of signals linked with the random contribution:
$$\sigma_R^2 = \frac{1}{N^2}\sum_{i=1}^N\sum_{j=1}^N  \left < \left |k^R_{ij} \right |^2 \right >\mbox{.}$$
The trace of $\mathbf{K}\mathbf{K^{\dag}}$ can be also expressed as
\begin{eqnarray*}
\mbox{Trace}\left [ \mathbf{K}\mathbf{K}^{\dag}\right ]&= & \sum_{i=1}^N\sum_{j=1}^N \left |k_{ij} \right |^2 \\
&= & \sum_{i=1}^N\sum_{j=1}^N \left (k^T_{ij}+k^R_{ij} \right) \left (k^{T*}_{ij}+k^{R*}_{ij} \right) \\
&= & \underbrace{ \sum_{i=1}^N\sum_{j=1}^N \left |k^T_{ij} \right |^2}_{N^2\sigma_T^2}  + \underbrace {\sum_{i=1}^N\sum_{j=1}^N \left |k^R_{ij} \right |^2}_{N^2 \sigma_R^2} + 2  \sum_{i=1}^N\sum_{j=1}^N  \Re \left [k^T_{ij} k^{R*}_{ij}\right]
\end{eqnarray*}
Assuming that $N\sigma_T>>\sigma_R$ and $N\sigma_R>>\sigma_T$ (assumptions verified \textit{a posteriori} with Eq.\ref{eqn:detection_RMT_2}), we can neglect the third sum because $\left< \Re \left[k^T_{ij} k^{R*}_{ij}\right] \right >=0$ and $\mbox{std}\left [  \sum_{i=1}^N\sum_{j=1}^N  \Re \left[k^T_{ij} k^{R*}_{ij}\right]\right]=N\sigma_T\sigma_R << N^2\sigma_T^2\mbox{, }N^2\sigma_R^2 $. And we finally obtain
\begin{equation}
\mbox{Trace}\left [ \mathbf{K}\mathbf{K}^{\dag}\right ]= N^2 \left (\sigma_T^2+ \sigma_R^2 \right)
\label{eqn:trace5}
\end{equation}
From Eq.\ref{eqn:trace4} and Eq.\ref{eqn:trace5}, we can deduce an expression for the quadratic mean of the singular values
\begin{equation}
\label{eqn:trace6}
 \sqrt { \frac{1}{N}\sum_{p=1} \lambda_p^2}=\sqrt{N\left ( \sigma_T^2+\sigma_R^2 \right)}
\end{equation}

\end{document}